\documentclass[12pt]{iopart}

\usepackage{iopams}  
\usepackage{graphicx} % Required for including images
\graphicspath{{figures/}} % Location of the graphics files
\usepackage{booktabs} % Top and bottom rules for table
\usepackage[font=small,labelfont=bf]{caption} % Required for specifying captions to tables and figures

\expandafter\let\csname equation*\endcsname\relax
\expandafter\let\csname endequation*\endcsname\relax

\usepackage{amsmath}
\usepackage{amsfonts, amsthm, amssymb} % , amsmath For math fonts, symbols and environments
\usepackage{subcaption}
\usepackage{dcolumn}
\usepackage{xcolor}
\usepackage{hyperref}
\usepackage[normalem]{ulem}

\usepackage{tikz}
\usetikzlibrary{shapes.geometric, arrows.meta, fit}
\usetikzlibrary{positioning}

\definecolor{darkperiwinkle}{RGB}{153 153, 192}
\definecolor{brightorange}{RGB}{254, 80, 0}
\definecolor{carolinablue}{RGB}{153,186,221}
\definecolor{cornflowerblue}{RGB}{100,149,237}

\newcommand{\todo}[1]{\textcolor{red}{#1}}

\usepackage[compress]{cite}
\usepackage{orcidlink}

\begin{document}
\vspace*{-3.5cm}

\title[AsterX: GPU-accelerated GRMHD for dynamical spacetimes]{AsterX: a new open-source GPU-accelerated GRMHD code for dynamical spacetimes}

\author{
Jay V. Kalinani $^{1, \ast} \orcidlink{0000-0002-2945-1142}$, 
Liwei Ji $^{1} \orcidlink{0000-0001-9008-0267}$, 
Lorenzo Ennoggi $^{1} \orcidlink{0000-0002-2771-5765}$,
Federico G. Lopez Armengol $^{1} \orcidlink{0000-0002-4882-5672}$,
Lucas Timotheo Sanches $^{2} \orcidlink{0000-0001-6764-1812}$,
Bing-Jyun Tsao $^{3} \orcidlink{0000-0003-4614-0378}$,
Steven R Brandt $^{4} \orcidlink{0000-0002-7979-2906}$,
Manuela Campanelli $^{1} \orcidlink{0000-0002-8659-6591}$, 
Riccardo Ciolfi $^{5,6} \orcidlink{0000-0003-3140-8933}$, 
Bruno Giacomazzo $^{7,8,9} \orcidlink{0000-0002-6947-4023}$,
Roland Haas $^{10,11} \orcidlink{0000-0003-1424-6178}$,
Erik Schnetter $^{12,13,4} \orcidlink{0000-0002-4518-9017}$,
Yosef Zlochower $^{1} \orcidlink{0000-0002-7541-6612}$
}

\address{
$^{1}$ Center for Computational Relativity and Gravitation, \& School of Mathematical Sciences,
Rochester Institute of Technology, 85 Lomb Memorial Drive, Rochester, New York 14623, USA
\\$^{2}$ Programa de Engenharia de Sistemas e Computação (PESC/COPPE), Universidade Federal do Rio de Janeiro (UFRJ), Avenida Horácio Macedo 2030, Centro de Tecnologia, Bloco H, sala 319. Rio De Janeiro - RJ, Brazil 21941-914.
\\$^{3}$ Center for Gravitational Physics, Department of Physics, The University of Texas at Austin, Austin, TX 78712, U.S.A
\\$^{4}$ Center for Computation \& Technology, Louisiana State University, Baton Rouge, LA, United States of America
\\$^{5}$ INAF, Osservatorio Astronomico di Padova, Vicolo dell’Osservatorio 5, I-35122 Padova, Italy
\\$^{6}$ INFN, Sezione di Padova, Via Francesco Marzolo 8, I-35131 Padova, Italy
\\$^{7}$ Dipartimento di Fisica G. Occhialini, Universit\`a di Milano-Bicocca, Piazza della Scienza 3, I-20126 Milano, Italy
\\$^{8}$ INFN, Sezione di Milano-Bicocca, Piazza della Scienza 3, I-20126 Milano, Italy
\\$^{9}$ INAF, Osservatorio Astronomico di Brera, Via E. Bianchi 46, I-23807 Merate, Italy
\\$^{10}$ National Center for Supercomputing applications, University of Illinois, 1205 W Clark St, Urbana, IL, United States of America
\\$^{11}$ Department of Physics, University of Illinois, 1110 West Green St, Urbana, IL, United States of America
\\$^{12}$ Perimeter Institute for Theoretical Physics, Waterloo, Ontario, Canada
\\$^{13}$ Department of Physics and Astronomy, University of Waterloo, Waterloo, Ontario, Canada}
\ead{$^{\ast}$\texttt{jaykalinani@gmail.com}}
%\vspace{10pt}

\iffalse
\begin{abstract}
%Abstract goes here
The Einstein Toolkit has, for roughly three decades, been among the most widely used codes by students around the world to study numerical relativity. Until fairly recently, it has not been possible to use the Einstein Toolkit with GPUs, thus limiting the code's usefulness on modern architectures. With the public release of the CarpetX framework in 2023, this situation has begun to change. In this paper, we introduce AsterX, a set of general purpose capabilities for relativistic hydrodynamics within the Einstein Toolkit. Here we discuss its scaling properties, tests that have been applied, and describe future work that remains to be done.
\end{abstract}
\fi

\begin{abstract}
We present \texttt{AsterX}, a novel open-source, modular, GPU-accelerated, fully general relativistic magnetohydrodynamic (GRMHD) code designed for dynamic spacetimes in 3D Cartesian coordinates, and tailored for exascale computing. We utilize block-structured adaptive mesh refinement (AMR) through \texttt{CarpetX}, the new driver for the \texttt{Einstein Toolkit}, which is built on \texttt{AMReX}, a software framework for massively parallel applications. \texttt{AsterX} employs the Valencia formulation for GRMHD, coupled with the `Z4c' formalism for spacetime evolution, while incorporating high resolution shock capturing schemes to accurately handle the hydrodynamics. \texttt{AsterX} has undergone rigorous testing in both static and dynamic spacetime, demonstrating remarkable accuracy and agreement with other codes in literature. Using subcycling in time, we find an overall performance gain of factor 2.5 to 4.5. Benchmarking the code through scaling tests on OLCF's Frontier supercomputer, we demonstrate a weak scaling efficiency of about 67\%-77\% on 4096 nodes compared to an 8-node performance.

\end{abstract}
%\noindent{\it Keywords}: general relativity, YYYYYYYY, ZZZZZZZZZ
%\noindent{\submitto{\CQG}}
%\submitto{XXX}
\maketitle

\section{Introduction}
\label{Intro}

Recent years have witnessed remarkable strides in multimessenger astrophysics, driven by technological advancements and collaborative efforts among global observatories. This progress was kick-started by the breakthrough detection of gravitational waves (GWs) from a binary neutron star (BNS) merger event `GW170817' \cite{LVC:BNSDetection}, accompanied by rich variety of electromagnetic (EM) counterparts including a short gamma-ray burst (SGRB) \cite{LVC-GRB} and an ultraviolet/optical/infrared transient consistent with a radioactively powered kilonova \cite{Pian2017} (GRB\,170817A and AT2017gfo, respectively), flooding the entire EM spectrum from gamma-rays to radio \cite{LVC-MMA}. The comprehensive data collected from this single event provided critical insights into the origin of heavy elements through r-process nucleosynthesis \cite{Arcavi2017,Kasen2017,Pian2017}, the nature of short gamma-ray bursts \cite{LVC-GRB,Mooley2018b,Lazzati2018,Ghirlanda2019}, and the behavior of matter under extreme conditions \cite{Margalit2017,LVC:EOSPaper:2018,Raithel2019,Matt2021}. Undoubtedly, the plethora of science extracted from this event has been unprecedented. Subsequently, a number of compact binary mergers involving at least one neutron star have been detected through GWs \cite{abbott2020gw190425, abbott2021observation, ligo2024observation}, making such events a cornerstone of the field. In the next decade or two, next-generation ground-based detectors (3G) ~\cite{Punturo:2010zza,Ballmer:2022uxx} are expected to expand our ability to observe these systems. On much larger scales instead, accreting supermassive binary black holes (SMBBHs) continue to be as promising multimessenger sources due to their strong emission of both GWs and EM radiation. While low-frequency GWs have already been detected by pulsar timing arrays \cite{agazie2023nanograv} and are anticipated to be detected by the Laser Interferometer Space Antenna (LISA) \cite{amaro2017laser, amaro2023astrophysics}, discovering SMBBH mergers in this data would offer profound insights for astrophysics and cosmology (e.g., \cite{Bogdanovic2022}). Additionally, the combination of EM and GW data will allow for a more comprehensive understanding of the environments surrounding SMBBH mergers and provide unique cosmological tests, such as measuring the Universe's expansion history. 

General relativistic magnetohydrodynamic (GRMHD) simulations remain an indispensable tool for probing the intricate dynamics of relativistic plasmas in the vicinity of such compact astrophysical sources. For instance, in recent years, various GRMHD simulations of BNS and black hole-neutron star (BHNS) mergers have demonstrated that magnetic fields play a critical role in launching relativistic jets, which can act as a pre-cursor or potentially power short gamma-ray bursts (SGRBs) \cite{Ruiz2016, Ciolfi2019, Ciolfi2020a, Moesta2020, Sun2022, combi2023jets, Most2023, Kiuchi:2023obe}. Additionally, they can have a profound impact on matter ejection, contributing to the kilonova production \cite{fernandez2019long, CiolfiKalinani2020,curtis2023r, combi2023grmhd}. On the other hand, for accurate modelling of the gas dynamics in SMBBH merger simulations, magnetic fields prove to be fundamental in driving turbulence and accretion processes from the circumbinary disks (CBD) \cite{Noble:2012xz,Lopez_Armengol_2021, Noble:2021vfg} to the minidisks surrounding each SMBH \cite{Bowen:2016mci,Bowen:2017oot, Bowen:2019ddu, Combi:2021xeh, Avara2024}, as well as in jet generation \cite{Combi:2021xeh, Gutierrez:2021png,gutierrez2024nonthermal, Ennoggi2024}. 

As the computational demands of these multi-physics simulations are substantial, they are typically limited in time and/or length scales. Existing state-of-the-art GRMHD codes such as WhiskyMHD \cite{giacomazzo2007whiskymhd, giacomazzo2011accurate}, GRHydro \cite{Moesta2014}, HARM3D \cite{gammie2003harm, noble2006primitive, Berthier2021}, IllinoisGRMHD~\cite{Etienne:2015cea,Werneck+2022}, MHDuet~\cite{palenzuela2021simflowny, palenzuela2022large, izquierdo2024large}, Spritz \cite{cipolletta2020spritz, cipolletta2021spritz, kalinani2022implementing}, GRAthena++ \cite{cook2023gr}, employed to run such simulations, are designed to work on central processing unit (CPU) architectures that are not efficient enough for massive parallel processing. To tackle these challenges, modern GRMHD codes are currently being redesigned directly or being build on infrastructures that support graphical processing unit (GPU) architectures such as \texttt{Parthenon} \cite{Grete2023} (based instead on \texttt{Kokkos} library \cite{CarterEdwards20143202}),  and \texttt{AMReX} \cite{Zhang2019amrex, Zhang2021amrex} which also provide block structured adaptive mesh refinement (AMR) capabilities. This is also driven by the revolutionary advancement in latest high-performance computing (HPC) supercomputers that employ GPUs to achieve exascale performance. Availing this advantage, the newest GPU-friendly (GR)MHD codes coupled with either static or dynamical spacetimes such as \texttt{K-Athena} \cite{grete2020k}, \texttt{H-AMR} \cite{Liska2020}, \texttt{GRaM-X} \cite{Shankar2023} have reported a performance speed up of a factor 2-10 in challenging simulations. 

In this paper, we present \texttt{AsterX}, a GPU-accelerated GRMHD code designed to work with dynamical spacetimes in Cartesian coordinates. \texttt{AsterX} is built on top of \texttt{CarpetX}, the new driver for the \texttt{Einstein Toolkit} \cite{etk2023} that provides AMR capability via the AMReX framework. Developed entirely in C++ from scratch as an open-source code \cite{AsterXGitHub}, \texttt{AsterX} incorporates numerous algorithms derived from the \texttt{Spritz} code \cite{cipolletta2020spritz, cipolletta2021spritz,kalinani2022implementing}. Like \texttt{Spritz}, it also adopts the Valencia formulation for GRMHD equations, and evolves the staggered version of the vector potential to preserve the divergenceless character of the magnetic field. At the moment, the code supplies only analytical equations of state (EOS), and support for tabulated microphysical EOSs is currently underway. For dynamical evolution, we utilize the spacetime solver based on the `Z4c' formulation \cite{Bernuzzi2010, Hilditch2013} of the Einstein's field equations. Here, we also present a series of extensive tests in 1D, 2D, and 3D that validate our implementation, and provide performance benchmarks conducted on OLCF's Frontier supercomputer, scaling up to 4096 nodes.

The paper is organized as follows. In Section \ref{basiceq}, we briefly describe the theoretical formulations on which our code is based. Numerical implementation of the adopted schemes are outlined in
Section \ref{numimp}. In Section \ref{tests}, we showcase the results of various tests performed in special and general relativistic regimes. Performance benchmarks are summarized in Section \ref{perf}. Finally, we present our conclusions and future outlook in Section \ref{concl}. Unless stated otherwise, we  adopt the space-like signature as $(-,+,+,+)$ and use geometric units $G=c=M_\odot=1$, and follow the standard Einstein's convention for the summation over repeated indices.

%%%%%%%%%%%%%%%%%%%%%%%%%%%%%%%%%%%%%%%%%%%%%%%%%%%%%%%%%%%%%%%%%%%%%%%%
\section{Background equations}
\label{basiceq}
In this Section, we briefly provide the theoretical background, describing the formulations adopted in \texttt{AsterX}. First, we focus on the \textit{3+1 foliation} technique adopted to arrive at numerical solutions of the Einstein field equations, followed by the \textit{3+1 formulation} of general relativistic magnetohydrodynamics (GRMHD) equations adopted for matter evolution.
\subsection{3+1 spacetime}
\label{spacetime}
The geometry of the 4-dimensional spacetime is governed by the Einstein's equations
\begin{equation}\label{einsteineqs}
{
    G_{\mu\nu} = 8\pi T_{\mu\nu} \ ,
}
\end{equation}
where $G_{\mu\nu}$ is the Einstein tensor, and $T_{\mu\nu}$ is the total stress-energy tensor. Our implementation employs the standard decomposition of spacetime into 3-dimensional space-like hypersurfaces associated with the line element of the form
\begin{equation}\label{lineelement}
{ 
    {\rm d}s^2  =  g_{\mu\nu}{\rm d}x^\mu {\rm d}x^\nu = -\alpha ^2 dt^2 + \gamma_{ij}({\rm d}x^i+ \beta^i{\rm d}t)({\rm d}x^j+ \beta^j{\rm d}t) \ ,
}
\end{equation}
where $\gamma_{ij}$ are the spatial components of the spacetime 4-metric $g_{\mu\nu}$,  $\alpha$ is the \textit{lapse} function and $\beta^i$ is the \textit{shift} vector. We then define $\textbf{n}$ as the unit 4-vector orthonormal to the hypersurfaces $\Sigma_t$, %\LJ{\st{such that it is proportional to the gradient of $t$ as $n=-\alpha \nabla t$,}}
which has the following components
\begin{equation}
\label{unitvec}
\eqalign{
    n^{\mu} = \frac{1}{\alpha} \left( 1, -\beta^{i} \right), \quad \quad 
    n_{\mu} = \left( -\alpha,0,0,0 \right).
}
\end{equation}
The extrinsic curvature is given by the Lie derivative of the three-metric in the $\textbf{n}$ direction as
\begin{equation}
%    K_{ij} = -2 \partial_t \gamma_{ij} + \beta^k \partial_k \gamma_{ij}
%               + \beta^k \partial_i \gamma_{kj} + \beta^k \partial_j \gamma_{ik}
K_{\mu\nu} = -\frac{1}{2}\mathcal{L}_\textbf{n}\gamma_{\mu\nu}
\end{equation}
For the evolution of spacetime, we consider the conformal decomposition of the Z4 formulation, commonly denoted as \textit{Z4c} \cite{Bernuzzi2010, Hilditch2013}. 
In the Z4c system, the Einstein equations are augmented with an additional 4-vector $Z^\mu$ such that in the extended evolution system, the Hamiltonian and momentum constraints are damped to zero.
The gauge conditions employed with this system include the $1+\log$ slicing and the $\Gamma$-driver shift. For convenience, the so-call ADM variables $\alpha$, $\beta^i$, $\gamma_{ij}$, and $K_{ij}$ are made available whenever the Z4c state vector is updated.
%When necessary, Christoffel symbols are calculated from the three-metric and its first derivatives.
Thus, \texttt{AsterX} is decoupled from details of the Z4c formulation. For further  details of this evolution system, we refer to \cite{Bernuzzi2010, Hilditch2013}. 

%%%%%%%%%%%%%%%%%%%%%%%%%%%%%%%%%%
\subsection{Valencia formulation} 
\label{val}
Analogous to the 3+1 decomposition of 4D spacetime, we require a 3+1 formulation to evolve the matter field equations involving magnetic fields. This Section discusses the 3+1 GRMHD formulation implemented in \texttt{AsterX}.

First, we define the 4D rest-mass density current and the energy-momentum tensor for a perfect fluid as
\begin{equation}
\label{energymom}
    J^\mu  = \rho u^\mu, \quad \quad
    T^{\mu \nu}  = (\rho + \rho \epsilon+p)u^\mu u^\nu + pg^{\mu\nu} = h\rho u^\mu u^\nu + pg^{\mu\nu} \ , 
\end{equation}
where $p$ is the gas pressure, $\rho$ stands for rest-mass density, $u^\mu$ is the four velocity of the fluid, $\epsilon$ is the specific internal energy, and $h=1+ \epsilon+ p/\rho$ denotes the specific relativistic enthalpy. The equation for the conservation of matter current density i.e., the continuity equation, and for the conservation of energy-momentum are written respectively as
\begin{equation}
\label{conseqs}
\nabla_\mu J^\mu = \nabla_\mu \left(\rho u^\mu\right) = 0 \ ,  \quad \quad \nabla_\mu T^{\mu \nu}  = \nabla_\mu \left[ h\rho u^\mu u^\nu + pg^{\mu\nu} \right] = 0  \ . 
\end{equation}
To include the effect of electric and magnetic fields in the GRHD formalism, we start with the Faraday electromagnetic tensor $F^{\mu\nu}$ and its dual ${}^*\!F^{\mu\nu} = (1/2)\eta^{\mu\nu\sigma\delta}F_{\sigma\delta}$, which are given by the expressions 
\begin{equation}
\label{faradaytensor}
    F^{\mu \nu} = \tilde{U}^{\mu} E^{\nu} - \tilde{U}^{\nu} E^{\mu} - \eta^{\mu \nu \lambda \delta} \tilde{U}_{\lambda} B_{\delta} \ ,
\end{equation}
\begin{equation}
\label{faradaydual}
    ^{*}F^{\mu \nu} = \frac{1}{2} \eta^{\mu \nu \lambda \delta} F_{\lambda \delta} = \tilde{U}^{\mu} B^{\nu} - \tilde{U}^{\nu} B^{\mu} - \eta^{\mu \nu \lambda \delta} \tilde{U}_{\lambda} E_{\delta} \ ,
\end{equation}
where $E^{\mu}$ is the electric field, $B^{\mu}$ stands for the magnetic field, $\tilde{U}^{\mu}$ a generic observer's four-velocity, and $\eta^{\mu \nu \lambda \delta} = \frac{1}{\sqrt{-g}} \left[ \mu \nu \lambda \delta \right]$ is the Levi-Civita pseudo-tensor representing the volume element. For the evolution of electromagnetic fields, we have the well-known Maxwell's equations
\begin{equation}
\label{maxwelleqs} 
\nabla_\nu {}^*\!F^{\mu\nu} = 0 \ , \quad \quad
\nabla_\nu F^{\mu\nu} = 4 \pi \mathcal{J}^\mu \ ,
\end{equation}
where $\mathcal{J}^\mu$ is the charge-current four-vector, expressed using Ohm's law as 
\begin{equation}
\label{currentvector}
    \mathcal{J}^\mu = q u^\mu + \sigma F^{\mu\nu}u_\nu \ ,
\end{equation}
where $q$ is the proper charge density and $\sigma$ represents the electrical conductivity. Under the assumption of ideal MHD, which considers a perfectly conducting fluid such that $\sigma \rightarrow \infty$ and $F^{\mu\nu} u_\nu=0$ (implying that the comoving observer measures no electric field), the electromagnetic tensor can be re-written solely in terms of the magnetic field $b$ measured in the comoving frame as
\begin{equation}
\label{faradaytensorsB}
    F^{\lambda \delta} = \eta ^{\mu \nu \lambda \delta} b_\mu u_\nu, \qquad \qquad {}^*\!F^{\mu\nu} = b^\mu u^\nu - b^\nu u^\mu = \frac{u^{\mu} B^{\nu} - u^{\nu} B^{\mu}}{W} \ ,
\end{equation}
whereas the Maxwell equations become
\begin{equation}
\label{maxwellB}
    \nabla_\nu {}^*\!F^{\mu\nu} = \frac{1}{\sqrt{-g}}\partial_\nu \left[\sqrt{-g} \left(b^\mu u^\nu - b^\nu u^\mu \right) \right] = 0 \ ,
\end{equation}
where $W = 1/\sqrt{1- v^i v_i}$ is the Lorentz factor and $b$ is the magnetic field measured by a comoving observer, which can then be written in terms of the magnetic field measured by an Eulerian observer $\vec{B}$ as
\begin{equation}
\label{eulerianB}
    b^0 = \frac{W B^i v_i}{\alpha}, \quad b^i = \frac{B^i + \alpha b^0 u^i}{W}, \quad b^2 \equiv b^\mu b_\mu = \frac{B^2 + \alpha^2 (b^0)^2}{W^2} \ ,
\end{equation}
with $B^2 \equiv B^i B_i$.
Considering $\tilde{B}^i \equiv \sqrt{\gamma}B^i$ where $\gamma$ is the determinant of $\gamma_{ij}$, and $\tilde{v}^i \equiv \alpha v^i - \beta^i$, Equation (\ref{maxwellB}) can be separated into time and spatial components, with the time component giving the \textit{divergence-free condition} as 
\begin{equation} 
\label{divB}
    \partial_i \tilde{B}^i =0 \ ,
\end{equation}
while the spatial component giving the magnetic-field \textit{induction equations} as
\begin{equation} 
\label{inductioneq}
    \partial_t \tilde{B}^i = \partial_j (\tilde{v}^i \tilde{B}^j - \tilde{v}^j \tilde{B}^i)  \ ,
\end{equation}
Including the magnetic field contribution, the stress-energy tensor can be redefined as
\begin{equation} 
\label{tmunu}
    T^{\mu \nu} = \left( \rho h + b^2\right) u^\mu u^\nu + \left( p + p_{\rm mag} \right) g^{\mu \nu} - b^\mu b^\nu \ ,
\end{equation}
where $p_\mathrm{mag} \equiv \frac{b^2}{2}$ is the magnetic pressure.
The set of Equations (\ref{conseqs}) can then be written in the following hyperbolic, first-order flux-conservative form
\begin{equation} 
\label{fluxconseqs}
    \frac{1}{\sqrt{-g}} \{ \partial_t [\sqrt{\gamma} \textbf{F}^0(\textbf{U})] + \partial_i [\sqrt{-g} \textbf{F}^{(i)}(\textbf{U})]\} = \textbf{S}(\textbf{U}) \ ,
\end{equation}
where $\textbf{F}^{(i)}(\textbf{U})$ stands for the flux-vector in the $i$-direction, $\textbf{S}(\textbf{U})$ are the source terms, and $\sqrt{-g} = \alpha\sqrt{\gamma}$. Except for the dependence on stress-energy tensor, the source terms depend only on the metric and its derivatives. In order to recast Equation (\ref{conseqs}) into a conservative form (\ref{fluxconseqs}), the \textit{primitive} magnetohydrodynamical variables $\textbf{U} \equiv (\rho,v^i,\epsilon, B)$ are mapped into a set of conserved variables $\textbf{F}^0(\textbf{U}) \equiv (D, S_i, \tau, \tilde{B})$ which are defined in the following way:
 
\begin{equation}
\label{grmhdeqs}
    \textbf{F}^0=\left(
    \begin{tabular}{ccc}
			$D$ \\
			$S_j$ \\
			$\tau$ \\
			$B^k$	
    \end{tabular}
    \right), \qquad \qquad \textbf{F}^i = \left(
    \begin{tabular}{ccc}
			$D\tilde{v}^i / \alpha$ \\
			$S_j \tilde{v}^i / \alpha + (p +b^2/2)\delta^i_j - b_jB^i/W$ \\
			$\tau \tilde{v}^i /\alpha + (p + b^2/2)v^i - \alpha b^0 B^i /W$ \\
			$B^k\tilde{v}^i / \alpha - B^i \tilde{v}^k /\alpha$
    \end{tabular}
    \right)  \ ,
\end{equation}
\begin{equation}
\label{grmhdsrc}
    \textbf{S} = \left(
    \begin{tabular}{ccc}
			$0$ \\
			$T^{\mu \nu} (\partial_\mu g_{\nu j} - \Gamma^\delta _{\nu \mu}g_{\delta j}) $ \\
			$\alpha (T^{\mu 0} \partial_\mu \ln \alpha - T^{\mu \nu} \Gamma^0_{\nu \mu})$ \\
			$0^k$
		\end{tabular}
		\right)	\ ,
\end{equation}
where $0^k = (0,0,0)^T$ and 
\begin{align}
\label{cons}
D & \equiv \rho W  \\
S_j & \equiv (\rho h + b^2)W^2v_j - \alpha b^0 b_j  \\
\tau & \equiv (\rho h + b^2)W^2 - (p+ b^2/2) - \alpha^2 (b^0)^2 - D \ . 
\end{align}
Along with Equations (\ref{grmhdeqs}-\ref{grmhdsrc}), we require an equation of state (EOS) which relates the pressure to the rest-mass density $\rho$ and to the specific internal energy $\epsilon$, in order to close the system of equations for hydrodynamics.

%%%%%%%%%%%%%%%%%%%%%%%%%%%%%%%%%%
\subsection{Electromagnetic gauge conditions} 
\label{emgauge} 

To ensure the divergence-free constraint (\ref{divB}), one common approach is to evolve the electromagnetic vector potential instead of magnetic field variables. To do so, we first express magnetic field $\vec{B}$ in terms of the vector potential $\vec{A}$. Considering $\nabla$ as a purely spatial operator, we can write $\vec{B} = \nabla \times \vec{A}.$
Taking the divergence of $\vec{B}$, we get
\begin{equation} 
\label{divB2}
    \nabla \cdot \vec{B} = \nabla \cdot ( \nabla \times \vec{A} ) = 0 \ .
\end{equation}
Therefore, by construction, evolving the vector potential $\vec{A}$ automatically satisfies (\ref{divB2}).

We now introduce the four-vector potential as $\mathcal{A}_{\nu} = n_{\nu} \Phi + A_{\nu},$
where $A_{\nu}$ is the purely spatial vector potential, whereas $\Phi$ stands for the scalar potential. The magnetic field, as measured by an Eulerian observer, can then be written as 
\begin{equation} 
\label{BEulerian}
    B^i = \epsilon^{ijk} \partial_j A_k \, ,
\end{equation}
and the induction equations (\ref{inductioneq}) can be re-written in terms of $A_i$ as
\begin{equation} 
\label{Ainductioneqs}
    \partial_t A_i = -E_i - \partial_i \left( \alpha \Phi - \beta^j A_j \right),
\end{equation}
where $\epsilon^{ijk} = n_{\nu} \epsilon^{\nu ijk}$ is the 3-dimensional spatial Levi-Civita tensor.

However, the gauge freedom admitted by Maxwell's equations renders the choice of the four-vector potential $\mathcal{A}^{\nu}$ as not unique, and allows setting a suitable gauge. 

\subsubsection{Algebraic gauge} \label{sec:ag}

The numerical codes that first employed vector potential evolution equations to perform GRMHD simulations considered using the \textit{algebraic} gauge~\cite{etienne2012relativistic,etienne2010relativistic}, where the scalar potential is set to $\Phi = \frac{1}{\alpha} \left( \beta^j A_j \right)$.
This choice leads to a much simplified version of (\ref{Ainductioneqs}), reducing it to $\partial_t A_i = - E_i$, and therefore, does not necessitate the evolution of the scalar potential $\Phi$.

\subsubsection{Lorenz gauge} \label{sec:lg}

The \textit{Lorenz} gauge is another choice that has been recently adopted in GRMHD simulations \cite{etienne2012relativistic}, which is based on the constraint $\nabla_{\nu} \mathcal{A}^{\nu} = 0$,
which serves as an advection equation for $\mathcal{A}$.
In order to impose this constraint, this gauge condition also requires solving the following evolution equation for the scalar potential
\begin{equation} 
\label{scalarpotentialevol}
    \partial_t \left( \sqrt{\gamma} \Phi \right) + \partial_i \left( \alpha \sqrt{\gamma} A^i - \sqrt{\gamma} \beta^i \Phi \right) = 0.
\end{equation}

The Lorenz gauge has proven to be advantageous over the algebraic gauge in GRMHD simulations which utilize adaptive mesh refinement (AMR), for e.g., BNS and NSBH merger simulations \cite{etienne2012relativistic}. The use of algebraic gauge in such simulations with AMR leads to generation of static gauge modes which then cause interpolation errors at the refinement boundaries, producing spurious magnetic fields near the boundary regions (see~\cite{etienne2012relativistic} for more details), rendering this gauge choice unsuitable for BNS, NSBH or SMBBH merger simulations, and more generally, for all simulations which involve magnetized matter crossing
refinement boundaries. 
A more robust gauge choice has been introduced in \cite{Farris2012} with the name of \textit{generalized Lorenz} gauge, which considers
\begin{equation} 
\label{genLorenz}
    \nabla_{\nu} \mathcal{A}^{\nu} = \xi n_\nu \mathcal{A}^{\nu} \, ,
\end{equation}
%
%where $\xi$ is a parameter that is typically set to be equal to $1.5/\Delta t_{\rm max}$, where $\Delta t_{\rm max}$ is the timestep of the coarsest refinement level \cite{etienne2015illinoisgrmhd}. 
where $\xi n_\nu \mathcal{A}^{\nu}$ acts like a damping term in the advection equation for $\mathcal{A}$. When employing this gauge choice, the evolution equation for the scalar potential becomes
\begin{equation} 
\label{scalarpotentialevol2}
    \partial_t \left( \sqrt{\gamma} \Phi \right) + \partial_i \left( \alpha \sqrt{\gamma} A^i - \sqrt{\gamma} \beta^i \Phi \right) = -\xi \alpha \sqrt{\gamma} \Phi \, .
\end{equation}
%
%%%%%%%%%%%%%%%%%%%%%%%%%%%%%%%%%%%%%%%%%%%%%%%%%%%%%%%%%%%%%%%%%%%%%%%%
\section{Numerical implementation}
\label{numimp}
%
%%%%%%%%%%%%%%%%%%%%%%%%%%%%%%%%%%
\subsection{Storage location for grid variables}
\label{vloc}
\begin{figure}[!ht]
\begin{center}
\includegraphics[width=0.7\linewidth]{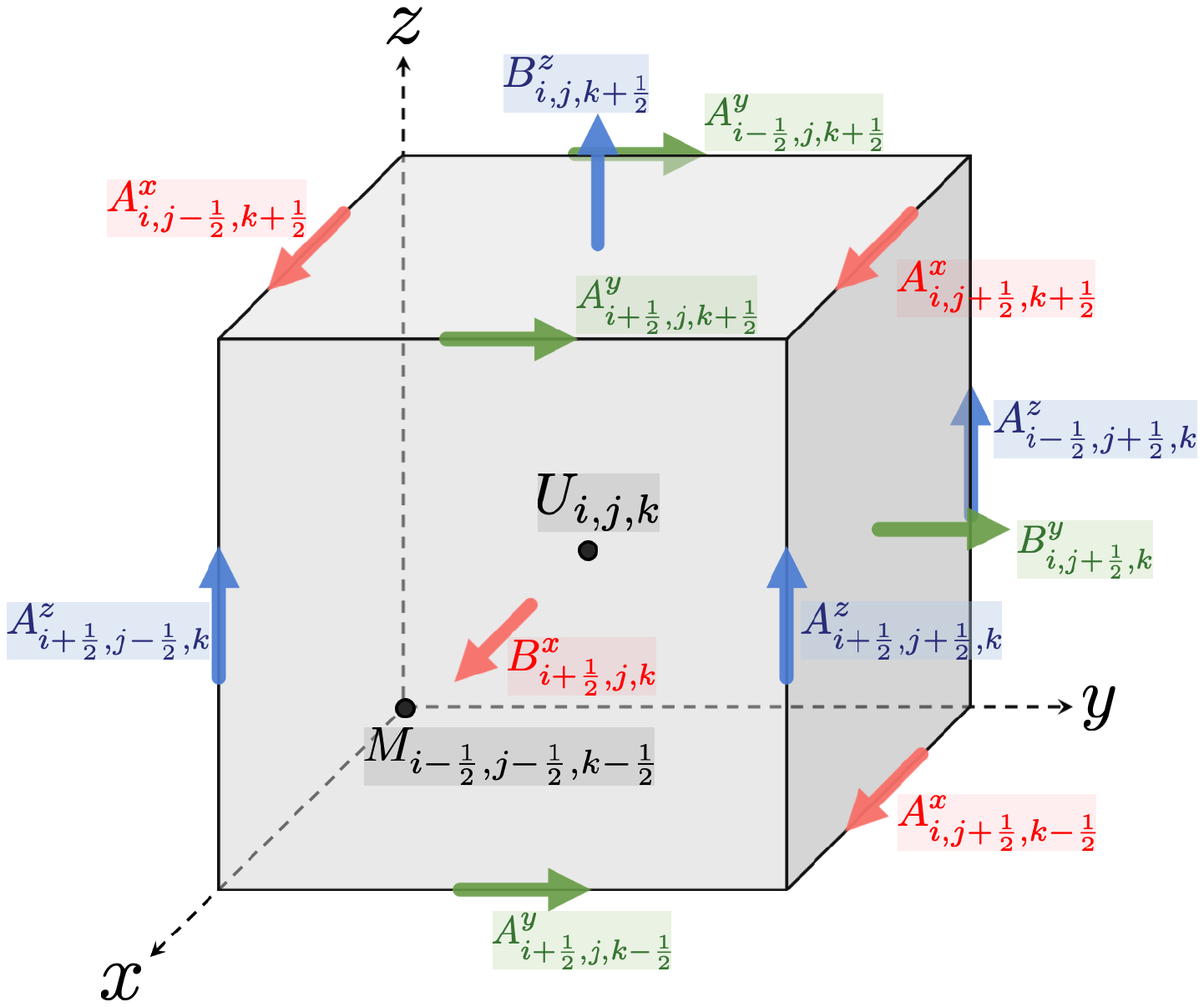}
\caption{Storage locations of different variables within a grid cell. Both conserved and primitive hydrodynamic quantities live at the cell center, denoted by $U_{i,j,k}$. Magnetic vector potential components $A_i$ are stored at the respective cell edges, whereas the magnetic field components $B^i$ and fluxes are at cell faces. The magnetic scalar potential as well as the spacetime variables are stored at cell vertices, represented by $M_{i-\frac{1}{2},j-\frac{1}{2},k-\frac{1}{2}}$. Figure adapted from \cite{cipolletta2020spritz}.}
\label{fig:cell}
\end{center}
\end{figure}

\texttt{CarpetX} allows for grid variables to be stored at centers, vertices, faces and edges of the grid cell. All the spacetime quantities including $\alpha$, $\beta^i$, $\gamma_{ij}$, and $K_{ij}$, as well as the stress-energy tensor $T_{\mu \nu}$ live at cell vertices. The hydrodynamic variables including primitives ($\rho$, p, $\epsilon$, $v^i$) as well as conservatives (D, $\tau$, $S_i$) are stored at cell centers. Magnetic vector potential components $A_i$ are staggered and are located at respective cell edges, taking the curl of which conveniently results in the magnetic field components $B^i$ at cell faces. Using the adjacent cell-face values, a linear interpolation is then employed to calculate the magnetic field components at the cell-centers, whenever necessary. The scalar potential $\Phi$ is instead stored at the cell vertex. Fluxes are also stored at cell faces. This is highlighted in Figure \ref{fig:cell}.
%
%%%%%%%%%%%%%%%%%%%%%%%%%%%%%%%%%%
\subsection{Spacetime evolution}
\label{stimp}
As mentioned in Section \ref{spacetime}, we use the \texttt{Z4c} module to evolve the spacetime. In our simulations, we employ the standard gauge and constraint damping parameters, specifically $\kappa_1=0.02$ and $\kappa_2=0.0$ for constraint damping, along with the lapse parameter $\mu_L=2/\alpha$, and shift parameters $\mu_S=1$ and $\eta=2$ \cite{Hilditch2013}.

Since all spacetime quantities are stored at grid vertices, which is different from the hydrodynamical quantities that are located at cell centers and fluxes at cell faces, we use fourth order Lagrange interpolation for interpolating the spacetime variables from vertices to cell centers or faces when evaluating the hydrodynamics sources and fluxes.

%%%%%%%%%%%%%%%%%%%%%%%%%%%%%%%%%%
\subsection{High resolution shock capturing schemes}
\label{hrsc}

Accurate and efficient treatment of shocks and discontinuities require state-of-the-art high-resolution shock capturing (HRSC) methods. Two main components of such schemes are (i) the reconstruction algorithms, employed to compute values of primitive variables at the cell interfaces, and (ii) the approximate Riemann solvers, used to compute the fluxes at these cell interfaces.

In \texttt{AsterX}, we have implemented the second order accurate total variation diminishing (TVD) MINMOD scheme \cite{Roe:1986cb}, as well as the third order accurate piecewise-parabolic method (PPM)\cite{Colella:1982ee, mccorquodale:11}. Higher order methods, i.e. WENO-Z \cite{borges2008wenoz}, and MP5 \cite{Suresh:1997am} have also been implemented, and currently are undergoing testing. All the schemes are made available through the \texttt{ReconX} module. Within this module, we also conduct internal unit tests for some of the aforementioned reconstruction algorithms to verify that their specific mathematical properties are satisfied. 

Approximate Riemann solvers implemented include Lax-Friedrichs (LxF) \cite{toro2013riemann} and Harten--Lax--van--Leer--Einfeldt (HLLE) \cite{harten1983upstream}. For HLLE, the numerical fluxes at cell interfaces are computed as
\begin{equation}
\label{HLLEfluxes}
\bi{F}^i = \frac{c_\mathrm{min} \bi{F}^i_\mathrm{r} + c_\mathrm{max} \bi{F}^i_\mathrm{l} - c_\mathrm{max}  c_\mathrm{min} \left( \bi{F}^0_\mathrm{r} - \bi{F}^0_\mathrm{l} \right)}{c_\mathrm{max} + c_\mathrm{min}} \, ,
\end{equation}
where the subscript r (l) denotes function computation at the right (left) side of the cell interface, and
$c_\mathrm{max} \equiv \max \left( 0, c_{+,l}, c_{+,r} \right)$,
$c_\mathrm{min} \equiv -\min \left( 0, c_{-,l}, c_{-,r} \right)$, where 
$c_{\pm,r}$ ($c_{\pm,l}$) are the right-going ($+$) and left-going ($-$) maximum characteristic wave speeds computed from the primitive variables $\bi{U}$ at the right (left) interface. Here, the computed eigenvalues are based on the solutions of a quartic equation, as described in detail for instance, in Section 3.1.1 of \cite{giacomazzo2007whiskymhd}.

Fluxes via LxF can instead be obtained by the expression 
\begin{equation}
\label{LxFfluxes}
\bi{F}^i = \frac{\bi{F}^i_\mathrm{r} + \bi{F}^i_\mathrm{l} - c_\mathrm{wave} \left( \bi{F}^0_\mathrm{r} - \bi{F}^0_\mathrm{l} \right)}{2},
\end{equation}
where $c_\mathrm{wave} = \max(c_\mathrm{max}, c_\mathrm{min})$~\cite{del2003efficient}. This scheme is more dissipative than HLLE, and can be useful in cases when dealing with strong shocks and interiors of black-hole horizons.
%
%%%%%%%%%%%%%%%%%%%%%%%%%%%%%%%%%%
\subsection{Electromagnetic field evolution}
\label{emevol}
In \texttt{AsterX}, we evolve the vector potential $A_i$ instead of the magnetic field variable  $B^i$, and then compute $B^i$ by taking the curl of the vector potential given by the Equation (\ref{BEulerian}). For the electromagnetic gauge, we have the choice to employ either the algebraic gauge or the generalized Lorenz gauge (\ref{genLorenz}). The latter is adopted by default in our simulations, unless stated otherwise.

When making use of the generalized Lorenz gauge, the damping term $\xi$ in Equation (\ref{scalarpotentialevol2}) is typically set to $1.5/\Delta t_{\rm max}$, where $\Delta t_{\rm max}$ is the time step of the coarsest refinement level, as also adopted in \cite{Cipolletta:2019:arXiv,etienne2015illinoisgrmhd}. 

Recasting the evolution equations of the vector potential and the densitized scalar potential ($\Psi_\mathrm{mhd} \equiv \sqrt{\gamma} \Phi$), the update terms for the right-hand side can be written as
\begin{align} 
\label{recastedAvec}
\partial_t A_i &= -E_i - \partial_i \left( G_A \right) \nonumber \\
& = -\epsilon_{ijk} \tilde{v}^j B^k - \partial_i \left( \alpha \frac{\Psi_\mathrm{mhd}}{\sqrt{\gamma}} - \beta^j A_j \right) \ ,
\end{align}
\begin{align} 
\label{recastedScalarPot}
\partial_t \Psi_\mathrm{mhd} &= -\partial_j \left( {F_{\Psi}}^j \right) - \xi \alpha \Psi_\mathrm{mhd} \nonumber \\
&= -\partial_j \left( \alpha \sqrt{\gamma} A^j - \beta^j \Psi_\mathrm{mhd} \right) - \xi \alpha \Psi_\mathrm{mhd} \ .
\end{align}

In the above equations, since some of the terms on the right-hand side are stored or staggered in different ways within a cell, they need to be interpolated at a common point. For solving Equation (\ref{recastedAvec}), we first interpolate $A_j$ to the vertices, where $\Psi_\mathrm{mhd}$ and the metric components are located. This allows us to compute $G_{A}$ at the vertices. We apply finite differencing with a stencil that computes the derivative of $G_{A}$ at cell edges, which gives the second term of the RHS.

To calculate the electric field in Equation (\ref{recastedAvec}) at cell edges, we have implemented two different schemes. The first approach uses the flux constrained transport (FCT) method~\cite{balsara1999staggered}, in which the electric field is computed from the magnetic field HLLE fluxes (see, for instance, Equation (23) of \cite{mignone2021systematic}). However, this scheme might result in generation of numerical instabilities in magnetic field evolution in simulations such as binary neutron star mergers. These instabilities can be curbed by adding an appropriate amount of Kreiss-Oliger dissipation to the magnetic field variables. Our second approach implements the upwind constrained transport (UCT) scheme as described in \cite{del2007echo, mignone2021systematic}. To compute, for instance, the $E_z$ component of the electric field on the edge $\left(i + \frac{1}{2}, j + \frac{1}{2}, k \right)$, we follow the steps summarized below:
\begin{enumerate}
	
	\item[(i)] Using Equation (57) of \cite{del2007echo}, the upwind transverse velocities are first computed, for example, on cell-face along $x$ direction, i.e. $\left(i + \frac{1}{2}, j, k \right)$ as
	\begin{equation} \label{uct1}
		\bar{v}^j = \frac{a^x_+ \tilde{v}^{jL} + a^x_- \tilde{v}^{jR}}{a^x_+ + a^x_-} \ , \qquad j=y,z \ .
	\end{equation}
	where L and R stand for left and right upwind states of $\tilde{v}^i$ along an interface, and $a^x_\pm$ is defined as 
	\begin{equation} \label{uct2}
		a^x_\pm = \rm{max} \{ 0, \pm \lambda^x_\pm (U^L), \pm \lambda^x_\pm (U^R) \} \ .
	\end{equation}
	where $\lambda^x_\pm$ are the characteristic wave speeds calculated at both left and right states. 
	
	\item[(ii)] Then, the quantities $\tilde{v}^x$, $\tilde{v}^y$, $B^x$, $B^y$ are reconstructed from faces $\left(i + \frac{1}{2}, j, k \right)$ and  $\left(i, j + \frac{1}{2}, k \right)$ to the edge $\left(i + \frac{1}{2}, j + \frac{1}{2}, k \right)$ using the PPM reconstruction method, following Equations (59) and (60) of \cite{del2007echo}.
	
	\item[(iii)] Finally, using the above expressions, the $E_z$ component of the electric field on the edge $\left(i + \frac{1}{2}, j + \frac{1}{2}, k \right)$ can be written using Equation (61) of \cite{del2007echo}, as
	\begin{align} \label{eq3.20}
		\begin{split}
			E_z = & - \frac{a^x_+ \bar{v}^{xL} B^{yL} + a^x_- \bar{v}^{xR} B^{yR} - a^x_+ a^x_- \left( B^{yR} - B^{yL} \right)}{a^x_+ + a^x_- }
			\\ 
			& + \frac{a^y_+ \bar{v}^{yL} B^{xL} + a^y_- \bar{v}^{yR} B^{xR} - a^y_+ a^y_- \left( B^{xR} - B^{xL} \right)}{a^y_+ + a^y_- }
		    \ .
		\end{split}
	\end{align}

\end{enumerate}
The other components $E_x$ and $E_y$ are also derived in the same way. 

For solving Equation (\ref{recastedScalarPot}), we only need to interpolate $A_j$ to the vertices, where $\Psi_\mathrm{mhd}$ and the metric components are located. ${F_{\Psi}}^j$ is then calculated at vertices. Using finite differencing with a stencil that computes the derivative of ${F_{\Psi}}^j$ at cell vertices, we obtain all the required RHS terms for Equation (\ref{recastedScalarPot}).

%%%%%%%%%%%%%%%%%%%%%%%%%%%%%%%%%%
\subsection{Primitive variables recovery} 
\label{c2p}
At each time-step during the evolution, we require values of the primitive variables $\textbf{U}$ in order to compute the numerical fluxes. However, since we evolve the conservative variables, the system of equations \eqref{cons} need to be inverted  to retrieve the primitives. This must be done numerically, and is one of the most delicate parts of the time evolution step. This process is referred to as \textit{primitive variables recovery} or \textit{conservative-to-primitives} (C2P) inversion.

\smallskip\noindent
In \texttt{AsterX}, we have implemented two such C2P methods:
\begin{itemize}
    \item \textbf{2D Noble et al. scheme} \cite{noble2006primitive, Siegel2018}. This method is based on the reduction of \eqref{cons} to a system of two equations and its subsequent inversion via a Newton-Raphson technique.
        First, we define
        \begin{equation}\label{Noble2D z}
            z\equiv\rho hW^2 \,,
        \end{equation}
        and use expression \eqref{eulerianB} for $b^0$, together with
        \begin{equation}
            b_i = \frac{B_i}{W} + \alpha b^0 v_i\,,
        \end{equation}
        to write \eqref{cons} as
        \begin{equation}
            \label{Conserved momentum Eulerian variables}
            S_i = -n_\mu T^\mu_{\phantom{\mu}i} = \alpha T^0_{\phantom{0}i} = \left(\rho h + b^2\right)W^2 v_i - \alpha b^0 b_i = \left(z + B^2\right)v_i - \left(B^jv_j\right)B_i\,.
        \end{equation}
        Next, contracting the above equation with $B^i$, we get
        \begin{equation}
            \label{BS Eulerian variables}
            B^jS_j = \rho h W^2\left(B^jv_j\right) = z\left(B^jv_j\right) \,.
        \end{equation}
       Finally, we use the above relation while contracting \eqref{Conserved momentum Eulerian variables} with $S^i$, and recast \eqref{cons} in terms of $z$ and $v^2$ to get
        \begin{subequations}
            \begin{align}
                \label{Noble2D momentum}
                v^2\left(B^2 + z\right)^2 - \frac{\left(B^iS_i\right)^2\left(B^2 + 2z\right)}{z^2} - S^2 &= 0 \\
                \label{Noble2D energy}
                \tau + D - \frac{B^2}{2}\left(1 + v^2\right) + \frac{\left(B^iS_i\right)^2}{2z^2} - z + p &=0\,,
            \end{align}
        \end{subequations}
        where $B^2\equiv B^iB_i$ and $S^2\equiv S^iS_i$. The above system is solved for the unknowns $z$ and $v^2$ via a Newton-Raphson technique, and the primitives are then retrieved in the following order
        \begin{subequations}
            \begin{align}
                W &= \frac{1}{\sqrt{1 - v^2}} \\
                \rho &= \frac{D}{W} \\
                v^i &= \frac{\gamma^{ij}S_j}{z + B^2} + \frac{\left(B^jS_j\right)B^i}{z\left(z + B^2\right)}\label{Noble2D v} \\
                \epsilon &= \frac{1}{\gamma}\left(\frac{z}{\rho W^2} - 1\right)\,,\label{Noble2D epsilon}
            \end{align}
        \end{subequations}
        where \eqref{Noble2D v} comes from \eqref{Conserved momentum Eulerian variables} and \eqref{BS Eulerian variables}, while \eqref{Noble2D epsilon} stems from \eqref{Noble2D z} together with relativistic specific enthalpy $h$, and pressure $p$ is derived from the evolution EOS.

    \item \textbf{1D Palenzuela et al. scheme} \cite{palenzuela2015effects, Siegel2018}. This method is based on the reduction of \eqref{cons} to a single equation, whose root is found via Brent's technique \cite{NumericalRecipes}. Since the root is bracketed and the algorithm does not attempt to estimate the derivatives of the function whose root is being searched for, this scheme is both more robust and more accurate than Newton-Raphson based techniques (though it is usually slower). The scheme works as follows:
    \begin{enumerate}
        \item starting from the conservative variables at the current timestep, define the quantities
        \begin{equation}\label{Palenzuela qrst}
            q\equiv\frac{\tau}{D},\quad r\equiv\frac{S^2}{D^2},\quad s\equiv\frac{B^2}{D},\quad t\equiv\frac{B^iS_i}{D^\frac{3}{2}}\;;
        \end{equation}
        \item compute $hW$ using the primitives at the previous timestep, and solve
        \begin{equation}\label{Palenzuela Brent}
            f\left(x\right) = x - \hat{h}\hat{W} = x - \Big( 1 + \hat{\epsilon} + \frac{p(\hat{\rho}, \hat{\epsilon}, \hat{Y_e})}{\hat{\rho}} \Big) \hat{W} \, ,
        \end{equation}
        via Brent's method to get the value of the unknown $x\equiv \rho h W^2 / \rho W = hW$ in between the bounds $1 + q - s < x < 2 + 2q -s$, where quatities with a hat in \eqref{Palenzuela Brent} are calculated at every iteration step from $x$;
        \item in terms of \eqref{Palenzuela qrst} and $x$ and using $v^2 = 1 - \frac{1}{W^2}$, equation \eqref{Noble2D momentum} can be recast as
        \begin{equation}
            \frac{1}{\hat{W}^2} = 1 - \frac{rx^2 + t^2\left(2x + s\right)}{x^2\left(x + s\right)^2}\;;
        \end{equation}
        \item compute $\hat{\rho} = \frac{D}{\hat{W}}$;
        \item again, make use of \eqref{Palenzuela qrst}, $x$ and $v^2 = 1 - \frac{1}{W^2}$ to recast \eqref{Noble2D energy} as
        \begin{equation}
            \hat{\epsilon} = - 1 + \frac{x}{\hat{W}}\left(1 - \hat{W}^2\right) + \hat{W}\left[1 + q - s + \frac{1}{2}\left(\frac{t^2}{x} + \frac{s}{\hat{W}^2}\right)\right]\;;
        \end{equation}
        \item using the evolution EOS, invert $\hat{\epsilon}$ to compute pressure $p$;
        \item noting that \eqref{Noble2D z} holds, and that $x$ is close to $hW$, the 3-velocity components can be obtained from \eqref{Noble2D v} setting $z = \rho Wx$.
    \end{enumerate}
\end{itemize}

Our C2Ps are made available through the \texttt{Con2PrimFactory} module.
%%%%%%%%%%%%%%%%%%%%%%%%%%%%%%%%%%
\subsection{Atmosphere}
\label{atmo}
One of the limitations of the state-of-the-art GRMHD codes is the difficulty to handle vacuum states. A standard approach imposes an artificial atmosphere, that requires setting a minimum density floor $\rho_{atm}$. In \texttt{AsterX}, during the evolution, if the conserved density `D' computed by the C2P falls below the floor density, i.e. $D < \sqrt{\gamma}\rho_{atm}$, then the primitive $\rho$ is reset to $\rho_{atm}$, whereas the pressure and specific internal energy are recomputed using a polytropic EOS. Moreover, the fluid’s three–velocity is set to zero. The conserved variables are then reset to atmospheric values using the modified primitives, via Equation (\ref{cons}). The magnetic fields are kept unchanged. For the tests presented in this paper, unless stated otherwise, we typically set $\rho_{atm} = 10^{-7} \rho_{0,\mathrm{max}}$, where $\rho_{0,\mathrm{max}}$ is the initial maximum value of the rest-mass density.

%%%%%%%%%%%%%%%%%%%%%%%%%%%%%%%%%%
\subsection{Equation of state}
\label{eos}
In order to close the system of equations \eqref{grmhdeqs}-\eqref{grmhdsrc}, an \textit{equation of state} (EOS) encoding the thermodynamic properties of the gas under study is needed. To this end, we couple \texttt{AsterX} to \texttt{EOSX}, a new Cactus module we developed that currently implements the following two analytical EOS
\begin{itemize}
    \item \textbf{ideal-fluid:}
        \begin{equation}
            P\left(\rho, \epsilon\right)=\left(\Gamma - 1\right)\rho\epsilon\;,
        \end{equation}
        where $\Gamma$ is the adiabatic index;
    \item \textbf{polytropic:}
        \begin{align}
            P\left(\rho\right) &= K\rho^\Gamma \, , \\
            \epsilon\left(\rho\right) &= \frac{K\rho^{\Gamma - 1}}{\Gamma - 1} = \frac{P\left(\rho\right)}{\left(\Gamma - 1\right)\rho} \, ,
        \end{align}
        where $K$ is the polytropic constant with $n=\frac{1}{\Gamma-1}$ as the polytropic index.
\end{itemize}
It is our intention, in future work, to render \texttt{EOSX} capable of supporting finite-temperature, tabulated EOS in the form $P\left(\rho, T, Y_e\right)$ and/or $P\left(\rho, \epsilon, Y_e\right)$.

%%%%%%%%%%%%%%%%%%%%%%%%%%%%%%%%%%
\subsection{Boundary conditions}
\label{bc}
\texttt{CarpetX} provides several types of boundary conditions that can be applied to the numerical domain, namely, `Dirichlet', `Neumann', `Robin', `linear extrapolation' and `none'. A Dirichlet boundary condition sets a fixed value of the function, whereas the Neumann boundary condition imposes a fixed value of the function derivative at the boundary. The Robin boundary condition is a mix of Dirichlet and Neumann conditions, specifying a linear combination of the function value and its derivative at the boundary. There is also support for periodic and reflection (symmetry) boundary conditions. Radiative boundary conditions are also made available via the \texttt{NewRadX} module. 
%%%%%%%%%%%%%%%%%%%%%%%%%%%%%%%%%%
\subsection{Adaptive mesh refinement}
\label{amr}

\texttt{CarpetX} offers the capability of adaptive mesh refinement through the \texttt{AMReX} framework \cite{Zhang2021amrex}. Support for fixed mesh refinement is provided by the \texttt{BoxInBox} module, which allows for constructing a hierarchical grid structure with nested rectangular boxes. Whereas, the dynamic (non-fixed) block-structured AMR enables higher refinement in localized regions of interest while maintaining a coarser resolution elsewhere, thus optimizing computational resources and enhancing accuracy. 
In the latter approach, once we have the solution over a rectangular grid, cells that require additional refinement can be identified through user-based criteria, which are then covered with a set of rectangular grids or blocks of higher resolution. \texttt{AMReX} follows the strategy of \cite{Berger1991} to determine the most efficient patch layout to cover the cells that have been tagged for refinement (see \cite{Zhang2021amrex} for more details).
In \texttt{AsterX}, we have implemented algorithms that compute either the gradient, first derivative, second derivative, or second derivative norm for one or more MHD variables, the results of which are compared to a user-defined error threshold parameter, to decide cell-tagging for refinement. 

%%%%%%%%%%%%%%%%%%%%%%%%%%%%%%%%%%
\subsection{Subcycling}
\label{sc}
Our latest implementation in \texttt{CarpetX} includes support for subcycling in time, in which grid variables on different levels progress with different time step sizes, thus reducing the computational effort required. This implementation follows a more efficient approach than the one provided in the original Carpet code. Specifically, subcycling in Carpet employs a buffer region with a width four times the number of ghost zones to supply inter-level boundary data at AMR boundaries during an RK4 evolution. Additionally, the buffer zone interpolation is second-order accurate, resulting in loss of accuracy. The new subcycling algorithm in \texttt{CarpetX}, based on~\cite{Mongwane2015, McCorquodale2011}, does not require any additional buffer zones, and all interpolations are fully fourth-order accurate. While this method will be presented soon in our upcoming paper (Ji et al., in preparation), we show preliminary results with \texttt{AsterX} employing subcycling in Section~\ref{perf} (see also Figure~\ref{fig:weakscalingsub}).

%%%%%%%%%%%%%%%%%%%%%%%%%%%%%%%%%%
\subsection{Parallelism}
\label{par}

As based on the \texttt{CarpetX} driver, \texttt{AsterX} can run on both CPUs and GPUs, and for convenience, code that has been written for GPUs can still execute efficiently on CPUs. We have tested \texttt{AsterX} with both NVIDIA GPUs (CUDA) and AMD GPUs (ROCm). Intel GPUs (oneAPI) are in principle also supported, but this has not been tested yet. This functionality is provided by \texttt{AMReX}.

\texttt{CarpetX} provides several levels of parallelism:
\begin{description}

\item[SIMD.] SIMD vectorization uses special CPU instructions to execute the same operation on multiple (usually 4 or 8) data elements simultaneously. Under ideal conditions, executing a SIMD instruction takes the same amount of time as executing a scalar instruction, and a compute kernel might thus run several times faster with this optimization. Support for SIMD instructions is implemented via the \emph{nsimd} library \cite{NSIMD}. Future versions of the C++ standard will support SIMD vectorization directly. Unfortunately, SIMD vectorization requires certain smallish changes to the source code, and not all parts of the code have been vectorized. In practice, code that is limited by memory bandwidth instead of compute throughput does not usually benefit from SIMD vectorization.

\item[Multi-threading.] Multi-threading is also called ``shared-memory parallelism.'' Modern CPUs have many cores, and it is beneficial to have multiple cores share data structures (grid functions) because this reduces overhead introduced by ghost zones (see below) that would be necessary by distributed-memory parallelism. \texttt{AMReX} splits grid functions into tiles, and we assign tiles to threads for calculations.
%: Each call to a compute kernel will handle one tile. We only use multi-threading for CPUs because GPU have already several levels of parallelism.

\item[GPUs.] GPUs over multiple levels of parallelism. Using CUDA terminology, a GPU executes \emph{threads} which perform individual operations. Threads are grouped into \emph{warps} which execute identical operations on different data elements. Warps are finally grouped into \emph{blocks} that operate on a common data structure. \texttt{CarpetX} hides these levels from the programmer, who writes compute kernels that are only slightly more complex than a CPU-only compute kernel. As mentioned above, a GPU-enabled compute kernel still runs efficiently on a CPU.

\item[Message passing.] This is also called ``distributed-memory parallelism.'' To split a large calculation onto multiple GPUs or multiple compute nodes we use \emph{MPI} to communicate between independently running processes. \texttt{CarpetX} hides this complexity from the programmer, using \emph{ghost zones} to allow the evaluation of finite differencing stencils and reconstructing fluid states on cell faces. This is also implemented via \texttt{AMReX}.

\end{description}

Overall, the programmer implements compute kernels that look very similar to straightforward C++ code, surrounded by function calls to allow \texttt{CarpetX} to parallelize the code when iterating over tiles.

%%%%%%%%%%%%%%%%%%%%%%%%%%%%%%%%%%
\subsection{AsterX workflow}
\label{workflow}

\begin{figure}[hbt!]
\begin{center}
\includegraphics[width=0.7\linewidth]{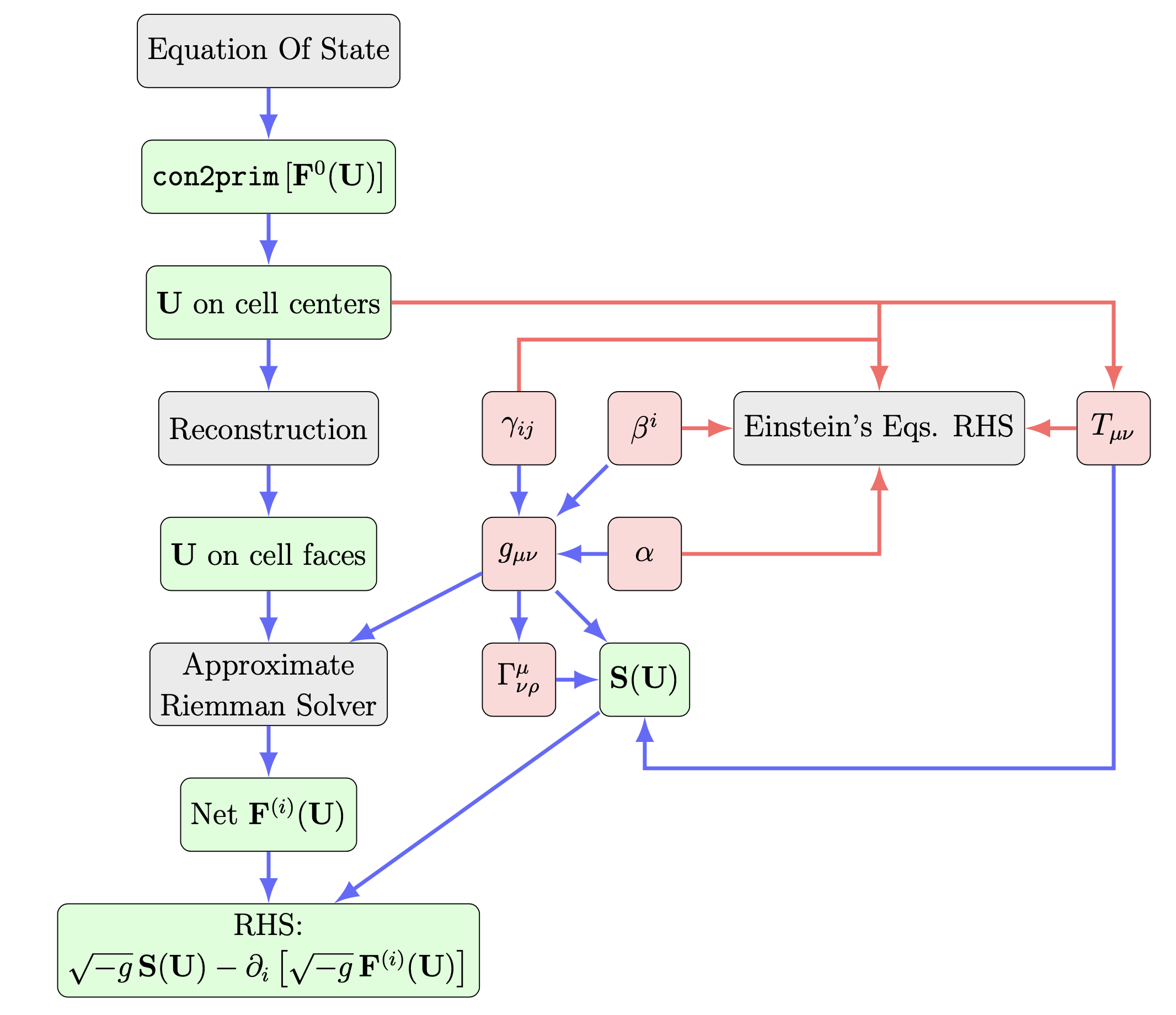}
\caption{Schematic representation of the AsterX workflow to compute the right-hand side of the conservative equations. See text for more details.}
\label{fig:asterx_workflow}
\end{center}
\end{figure}

To facilitate users' comprehension of the organizational structure of our code, Figure \ref{fig:asterx_workflow} illustrates an overview of the workflow within \texttt{AsterX} for computing the right-hand side of the GRMHD equations.

In the figure, elements outlined by green boxes denote GRMHD variables, red boxes represent Einstein's equations, while grey boxes show functional steps in the GRMHD workflow. Concurrently, blue arrows denote data flow for computing the GRMHD RHS, while those depicted by red boxes and arrows pertain to relevant segments of the spacetime evolution scheme.

The code is initialized with data encompassing both primitive hydrodynamical variables (refer to Section~\ref{val}) and spacetime quantities, including lapse, shift vector, spatial metric, and extrinsic curvature. Analytical expressions \ref{cons} are then employed to initialize the conserved variables. The code loops over the following steps until the designated final simulation time is reached: 

\begin{enumerate}
    \item Utilizing the conserved variables from the previous step and an appropriate EOS, \texttt{con2prim} is called to obtain cell averages of the conserved variables.
    \item Cell averages of the conserved variables are reconstructed on cell faces while also being used in computing the energy momentum tensor of the evolution.
    \item The Riemman problem on each cell face is solved via an approximate Riemman solver in order to obtain the numerical flux between cells. This process also requires the 3-metric along with lapse and shift (components of the four-metric) from the gravitational evolution part of the code.
    \item Once the fluxes have been computed, the 3-metric along with lapse and shift and their derivatives are used to compute source terms, finalizing the computation of the hydrodynamical RHS.
\end{enumerate}

%\include{workflow}

%%%%%%%%%%%%%%%%%%%%%%%%%%%%%%%%%%%%%%%%%%%%%%%%%%%%%%%%%%%%%%%%%%%%%%%%
\section{Tests}
\label{tests}
In this Section, we report on the results of our extensive testing,
including a number of 1–, 2– and 3–dimensional simulations. These simulations include
critical tests that have been already considered in the literature in several previous
papers (see, e.g., and references therein), allowing for a direct comparison
with other codes.

%%%%%%%%%%%%%%%%%%%%%%%%%%%%%%%%%%
%\subsection{1D tests}
%\label{1D}
%\todo{Jay}
%Here, we report results from the conventional 1-dimensional (1D) tests based on Balsara shock-tube problems as well as circularly polarized Alfv\'en wave, described in detail below.

\subsection{Balsara shocktube}
\label{shocktube1D}
\begin{figure}[hbt!]
\begin{center}
\includegraphics[width=\linewidth]{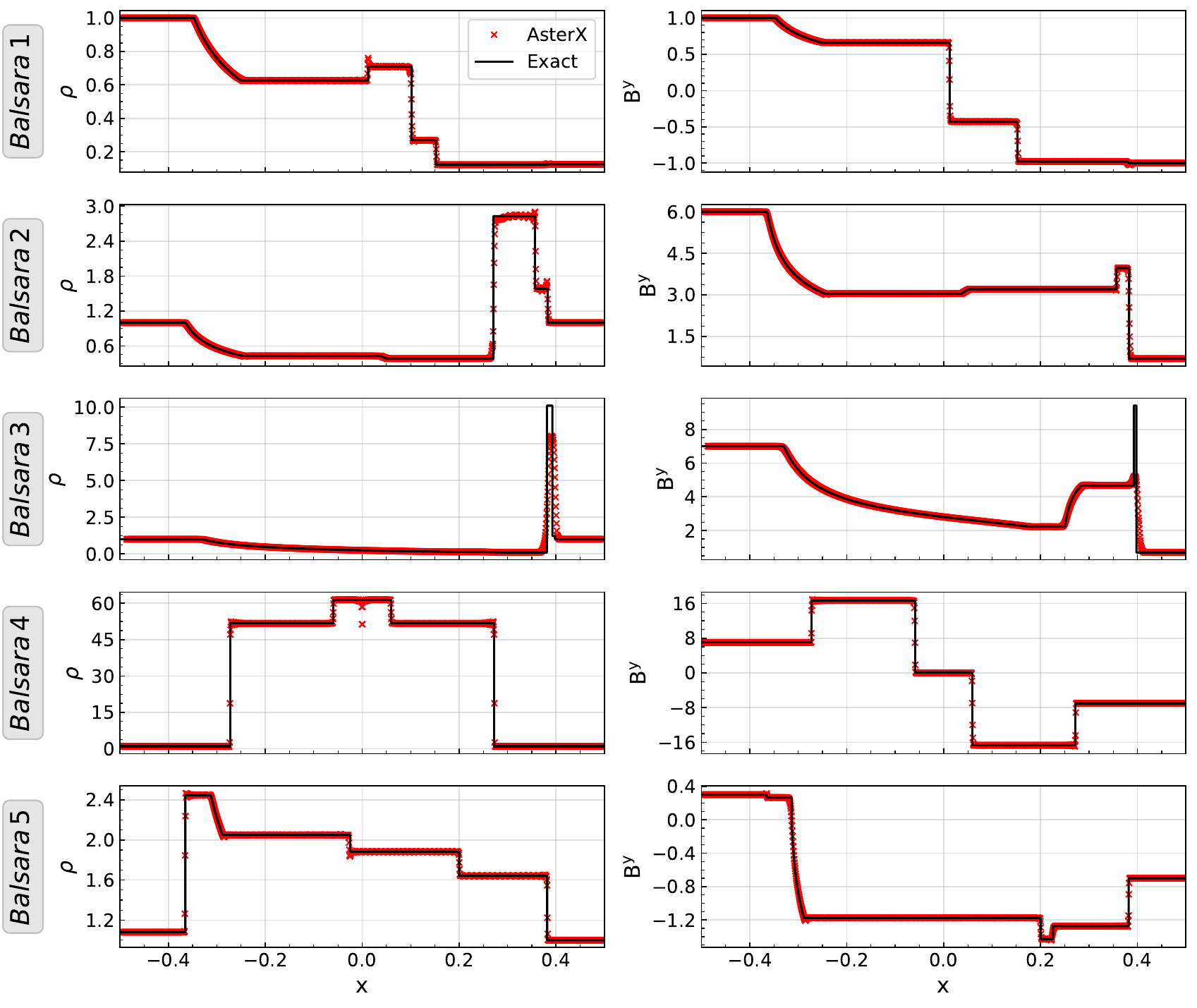}
\captionof{figure}{\label{fig:BalsaraST}Comparison of numerical results (red dots) and exact solutions (continuous black lines) for the suite of tests of \cite{balsara2001total}. Left and right columns show respectively the spatial distributions of the rest-mass density and the magnetic field component $B^y$ at the final time of the evolution. 
Here, the \texttt{Balsara 1}, \texttt{2}, \texttt{4} and \texttt{5} tests are performed with the third order PPM method. On the other hand, \texttt{Balsara 3} is performed with the second order MINMOD method because this test is the most demanding one due to the jump of four orders of magnitude in the initial pressure. This extreme change requires a slightly more dissipative method to succeed.}
\end{center}
\end{figure}

In order to check the correctness of the approximate Riemann solvers and other numerical schemes implemented in the code, the first round of tests we performed with \texttt{AsterX} are those involving Riemann problems.
In Figure~\ref{fig:BalsaraST}, we show the results for 1-dimensional (1D) relativistic shock-tube problems based on the testsuite of~\cite{balsara2001total}. Here, our numerical results of these tests are directly compared with the exact solutions computed via the code presented in~\cite{giacomazzo2006exact}. Initial data for such tests are described in Table~\ref{tab1}. 

\begin{table}[!ht]
	\footnotesize
	\begin{center}
		\begin{tabular}{@{}ccccccccccc}
		\toprule	
			Test: & \multicolumn{2}{c}{ \texttt{1} } & \multicolumn{2}{c}{ \texttt{2} } & \multicolumn{2}{c}{ \texttt{3} } & \multicolumn{2}{c}{ \texttt{4} } & \multicolumn{2}{c}{ \texttt{5} } \\
			\midrule
			& L & R & L & R & L & R & L & R & L & R \\
			 $\rho$ & 1.0 & 0.125 & 1.0   & 1.0 & 1.0       & 1.0  & 1.0     &  1.0      & 1.08 & 1.0   \\
			 $p$     & 1.0 & 0.1     & 30.0 & 1.0 & 1000.0 & 0.1  & 0.1     & 0.1      & 0.95 & 1.0   \\
			 $v_x$  & 0.0 & 0.0     & 0.0   & 0.0 & 0.0       & 0.0  & 0.999 & -0.999 & 0.4  & -0.45 \\
			 $v_y$  & 0.0 & 0.0     & 0.0   & 0.0 & 0.0       & 0.0  & 0.0     & 0.0      & 0.3  & -0.2   \\
			 $v_z$  & 0.0 & 0.0     & 0.0   & 0.0 & 0.0       & 0.0  & 0.0     & 0.0      & 0.2  & 0.2    \\
			 $B^x$  & 0.5 & 0.5    & 5.0    & 5.0 & 10.0    & 10.0 & 10.0   & 10.0    & 2.0 & 2.0     \\
		     $B^y$  & 1.0 & -1.0   & 6.0    & 0.7 & 7.0      & 0.7   & 7.0     & -7.0    & 0.3  & -0.7   \\
			 $B^z$  & 0.0 & 0.0    & 6.0    & 0.7 & 7.0      & 0.7   & 7.0     & -7.0    & 0.3  & 0.5     \\
			\bottomrule
		\end{tabular}\\
		\caption{\label{tab1}Initial data for \texttt{Balsara} relativistic shock tube tests.}
	\end{center}
\end{table}

For all tests, we employ an ideal fluid EOS, with $\Gamma = 2.0$ for test \texttt{Balsara 1} and $\Gamma = 5/3$ for the others. The final evolution time is considered as $t = 0.55$ for test \texttt{Balsara 5} and $t = 0.4$ for the others. We note that all tests show an excellent agreement between the numerical results and the exact solutions.

\subsection{Alfv\'en wave}
\begin{figure}[!ht] 
\begin{center}
\includegraphics[width=0.55\linewidth]{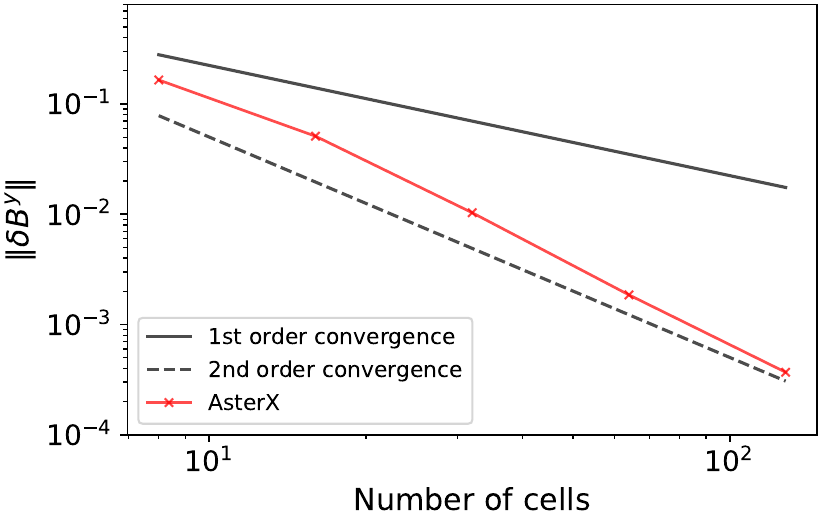}
\captionof{figure}{Results of the Alfv\'en wave test. Here, the L2 norm of error in the y-component of the magnetic field is illustrated (red solid line) as a function of the number of cells used along x-direction. Reference curves for 1st and 2nd order convergence are shown in black solid and dashed lines respectively. With the increase in resolution, our results show good agreement with the expected $\sim \!\!$ 2nd order convergence rate as well as with the ones reported in literature \cite{cipolletta2020spritz, Shankar2023}. }
\label{fig:AW}
\end{center}
\end{figure}

To determine the convergence order of our code, we perform the conventional Alfv\'en wave test. This involves advecting a circularly polarized Alfv\'en wave across the simulation domain. The solutions for such a wave are exact and do not contain shocks, making it an ideal candidate for this test. The initial conditions are adopted from~\cite{cipolletta2020spritz, Shankar2023}, where we set the wave amplitude $A_0=1.0$, the fluid rest-mass density $\rho=1.0$, fluid pressure $\rho=0.5$ and the Alfv\'en wave speed $v_A=0.5$. The velocity components are initialized as
\begin{equation}
\label{awvel}
    v^x = 0, \quad v^y = -v_A A_0 \cos{(kx)}, \quad v^z = -v_A A_0 \sin{(kx)} \ ,
\end{equation}
whereas the magnetic field components are initialized as
\begin{equation}
\label{awB}
    B^x = 1.0, \quad B^y = A_0 B^x \cos{(kx)}, \quad B^z = A_0 B^x \sin{(kx)} \ .
\end{equation}
The wave propagates along the x-direction with the wave vector $k=2\pi/L_x$ and the domain length $L_x=1$, where the grid along x extends from -0.5 to +0.5. We study convergence based on various different resolutions, varying the number of grid points along x as 8, 16, 32, 64 and 128.
For the evolution, we use the ideal gas EOS with $\Gamma=5/3$, MINMOD reconstruction, HLLE Riemann solver, RK2 for time-stepping with the Courant factor of 0.2, and periodic boundary conditions. The result of the L2 norm of the convergence error in the y-component of the magnetic field $B^y$ as a function of number of grid points is illustrated in Figure \ref{fig:AW}. 
In particular, we compute the L2 norm based on the initial and final wave profile after 1 full period of evolution. Here, we observe an overall 2nd order convergence rate, an expected outcome that has also been reported, for instance, in \cite{cipolletta2020spritz, Shankar2023} for the same test case.

%%%%%%%%%%%%%%%%%%%%%%%%%%%%%%%%%%
%\subsection{2D tests}
%\label{2D}
%\todo{Jay}
%magnetic rotor, magnetic loop advection and Kelvin-Helmholtz instability (KHI), that were carried out with AsterX. We describe each of them below in detail.

%%%%%%%%%%%%%%%%%%%%%%%%%%%%%%%%%%
\subsection{Cylindrical explosion}
\label{cylexp} 
The first test considers the evolution of a 2D cross-section of a dense, over-pressured cylinder in a uniformly magnetized environment. Following this blast wave allows a check of whether the code can correctly follow the shock front on the equatorial plane. The initial data is based on the setup described in \cite{Cipolletta:2019:arXiv}. For the cylinder, we consider
\begin{equation} \label{cylexp1}
    r_\mathrm{in}= 0.8, \; r_\mathrm{out}= 1.0, \; \rho_\mathrm{in} = 10^{-2}, \; p_\mathrm{in} = 1.0 , 
\end{equation}
while for the surrounding ambient medium, we set 
\begin{equation} \label{cylexp2}
	\rho_\mathrm{out} = 10^{-4}, \; p_\mathrm{out} = 3 \times 10^{-5}. 
\end{equation}
Here, the parameters $r_\mathrm{in}$ and $r_\mathrm{out}$ are used for smoothening the density profile (and similarly for the smoothening of the pressure profile) considered in~\cite{Cipolletta:2019:arXiv}, such that
\begin{align} \label{cylexp3}
	\rho(r) = 
	\begin{cases} 
	& \rho_\mathrm{in}  \, \qquad \qquad \qquad \qquad \qquad \qquad  {\rm for} \ r \leq  r_\mathrm{in}  \\
	& \exp \left[ \frac{(r_\mathrm{out} - r) \ln \rho_\mathrm{in} + (r-r_\mathrm{in}) \ln \rho_\mathrm{out}}{r_\mathrm{out} - r_\mathrm{in}}  \right] \, \ \quad {\rm for} \ r_\mathrm{in} < r <  r_\mathrm{out} \\ 
	& \rho_\mathrm{out}  \ \ \qquad \qquad \qquad \qquad \qquad \quad \  {\rm for} \ r \geq  r_\mathrm{out}
	\end{cases}
\end{align}
Initially, the fluid 3-velocity is set to zero while the magnetic field is kept uniform with $B^x = 0.1$ and $B^y = B^z = 0$. The grid domain spans the range $[-6,6]$ along x and y-axes containing $200 \times 200$ grid points. For the evolution, we use an ideal fluid EOS with adiabatic index $\Gamma = 4/3$ and set the CFL factor to 0.25. For solving the numerics, the MINMOD reconstruction method is employed along with the HLLE flux solver and the RK4 method for the time-step evolution. 

\begin{figure}[!ht] 
	\begin{center}
		\includegraphics[width=0.49\linewidth]{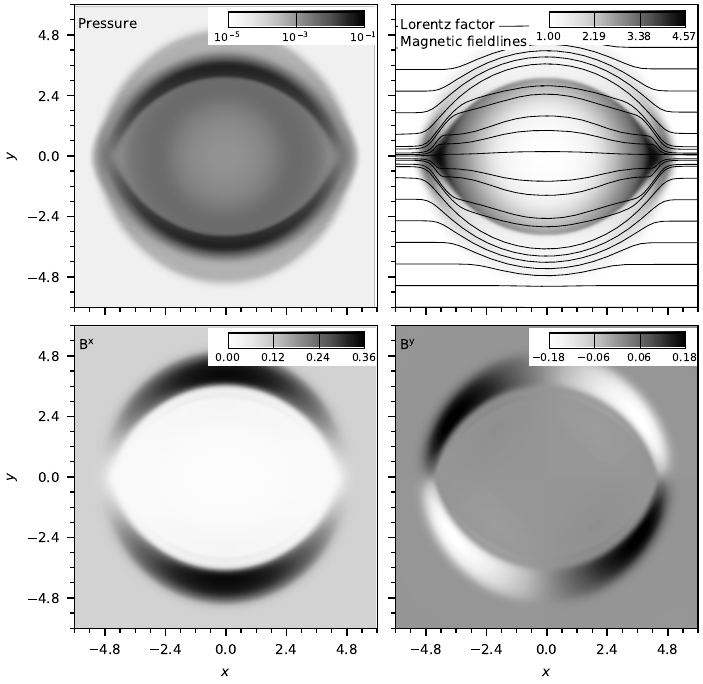}
            \includegraphics[width=0.49\linewidth]{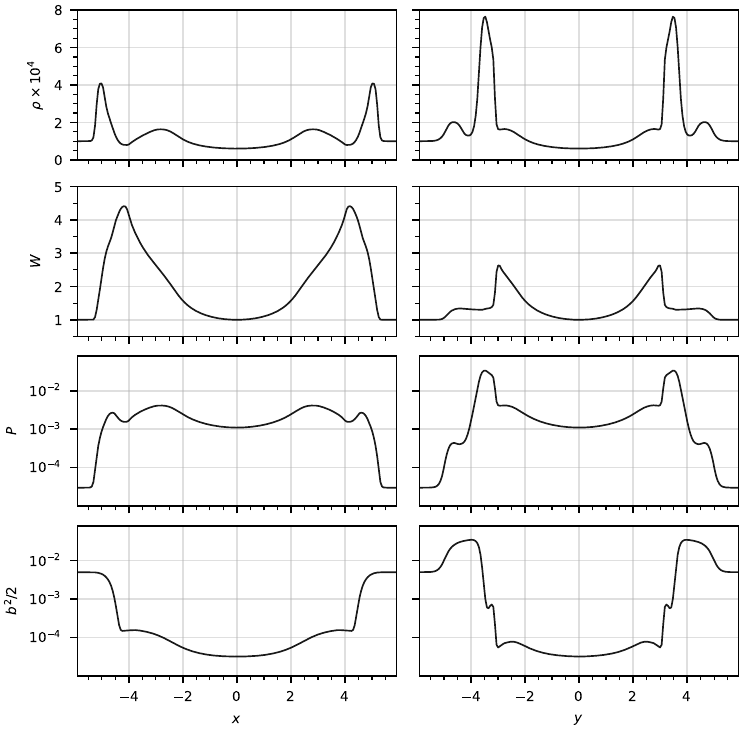}
            \captionof{figure}{Equatorial snapshots (left panel) and 1D slices along x and y-axis (right-panel) of the cylindrical explosion test at final time $t=4$. On the left, we have gas pressure $P$ (top-left), Lorentz factor $W$ together with magnetic field lines (top-right), and x-~and y-components of the magnetic field, $B^x$ (bottom-left) and $B^y$ (bottom-right). Right panel shows density $\rho$ (top row), lorentz factor $W$ (second row), gas pressure $P$ (third row) and magnetic pressure $b^2/2$ (bottom row). }
		\label{fig:CylBW}
	\end{center}
\end{figure}

In Figure~\ref{fig:CylBW}, the left panel illustrates the 2D blast wave profiles at the final time $t=4.0$, representing the distribution of gas pressure $p$, Lorentz factor $W$ (together with magnetic field lines), and the $x$-- and $y$--components of the magnetic field, $B^x$ and $B^y$. Profiles from 1D slices along x- and y-axis are instead shown in the right panel.
We find both 1D and 2D profiles to be in good agreement when compared to the results already presented in the literature \cite{Cipolletta:2019:arXiv}.

%%%%%%%%%%%%%%%%%%%%%%%%%%%%%%%%%%
\subsection{Magnetic rotor}
\label{magrot}

Another 2D test we consider is the magnetic cylindrical rotor, originally introduced for classic MHD in \cite{balsara1999staggered,toth2000b} and later employed also for relativistic MHD in \cite{etienne2010relativistic,del2003efficient}. This test consists of a dense, rapidly spinning fluid at the center, surrounded by a static ambient medium starting with a uniform magnetic field and pressure in the entire domain. For initial data, we set the radius of the inner rotating fluid to $r=0.1$, with inner rest-mass density $\rho_\mathrm{in} = 10.0$, uniform angular velocity $\Omega=9.95$, and therefore the maximum value of the fluid 3-velocity is $v_\mathrm{max} = 0.995$. In the outer static ambient medium, we set the rest-mass density to $\rho_\mathrm{out} = 1.0$. The initial magnetic field and gas pressure are set to $B^i=(1.0, 0, 0)$ and $p_{\mathrm{in}} = p_{\mathrm{out}} = 1.0$ respectively. The numerical domain is a $400 \times 400$ grid with $x$- and $y$-coordinates lying in range $[-0.5,0.5]$. Here, too, we set the CFL factor to 0.25, and consider an ideal fluid EOS with $\Gamma = 5/3$ for evolution. For the numerics, we use the MINMOD reconstruction, the HLLE flux solver, and the RK4 method for time-update.

\begin{figure}[!ht]
	\begin{center}
		\includegraphics[width=0.49\linewidth]{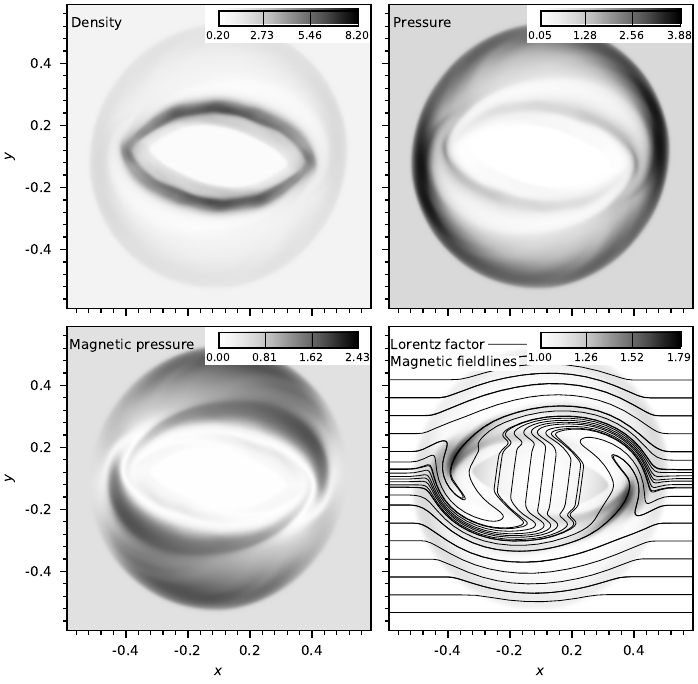}
            \includegraphics[width=0.49\linewidth]{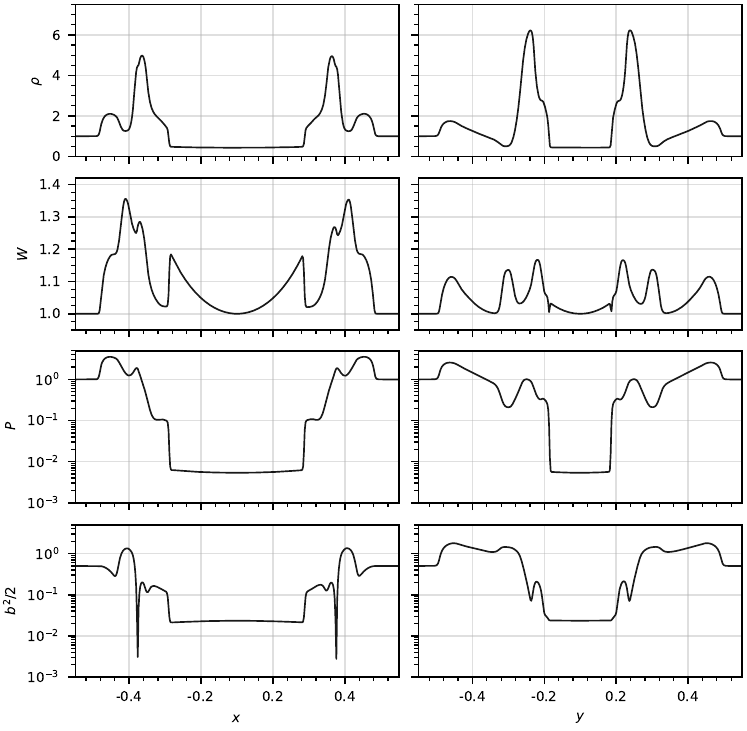}
		\caption{2D snapshots (left) and 1D slices along x- and y-axis (right) of the Magnetic rotor test for the final evolved time $t=0.4$. The left panel shows the density $\rho$ (top-left), gas pressure $p$ (top-right), magnetic pressure $p_\mathrm{mag}$ (bottom-left), and Lorentz factor $W$ together with the magnetic field lines (bottom-right). The right plot illustrates the density $\rho$, Lorentz factor $W$, gas pressure $p$ and magnetic pressure $b^2/2$ along the $x-$ and $y-$ axes respectively. The resolution here is $\Delta x = \Delta y = 0.0025$.}
		\label{fig:MagRot}
	\end{center}
\end{figure}

Figure~\ref{fig:MagRot} depicts the 2D profiles of density $\rho$, gas pressure $p$, magnetic pressure $p_\mathrm{mag}= b^2/2$, and Lorentz factor $W$ along with magnetic field lines at the final time $t=0.4$. 
The rotation of the cylinder leads to magnetic winding. This can be seen in the bottom-right panel of left Figure~\ref{fig:MagRot} where, near the central region, the field lines are twisted roughly by $\sim 90^\circ$. This twisting of field lines could eventually slow down the rotation of the cylinder. A decrease in $\rho$, $p$, and $p_\mathrm{mag}$ in the central region is also observed along with the formation of an oblate shell of higher density. Also for this test, the results are in good agreement with the ones in the literature \cite{Cipolletta:2019:arXiv,Shankar2023, Moesta2014,del2003efficient}.

For a quantitative check, we again take a 1D slice along $x=0$ and $y=0$ of the final rotor configuration at $t=0.4$, as shown in the right panel of \ref{fig:MagRot}. Comparing with the results presented in \cite{Cipolletta:2019:arXiv,Shankar2023}, we find the curves including the peak values in agreement with the literature.

\begin{figure}[!ht]
	\begin{center}
		\includegraphics[width=0.75\linewidth]{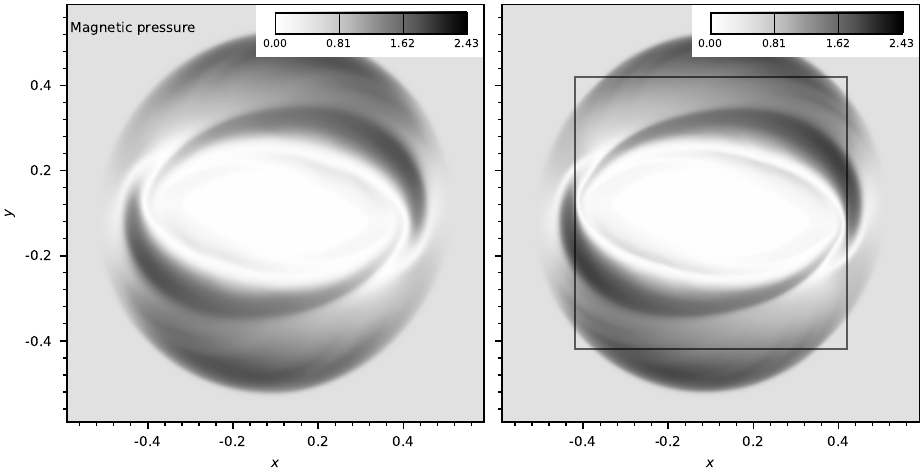}
		\caption{Magnetic rotor test with adaptive mesh refinement (AMR). For comparison, we show the final magnetic pressure configuration at t = 0.4 between the uniform grid test (left panel) and the AMR test (right panel) including a refined inner grid (black box) with twice the resolution. We find the two results to be agreement, without any spurious effects at the inner refinement boundary.}
		\label{fig:MagRotAMR}
	\end{center}
\end{figure}

To assess the AMR functionality, we conduct an additional simulation by introducing an inner refined grid spanning the x- and y-coordinates from [-0.32, +0.32], with half the grid spacing, utilizing the \texttt{BoxInBox} module. For prolongation at the inner refinement boundary, we interpolate polynomially in vertex centered directions for the spacetime variables, and conserve with third order accuracy along other directions for MHD variables, with a linear fallback in presence of shocks. Figure~\ref{fig:MagRotAMR} presents a comparison of the final magnetic pressure distribution with the results from the uniform grid test. The comparison shows no significant differences nor any noticeable effects at the boundaries between grids, confirming the correct working of our code with the AMR framework.

%%%%%%%%%%%%%%%%%%%%%%%%%%%%%%%%%%
\subsection{Magnetic loop advection}
\label{loopadv}

In this test, we simulate an advecting magnetic field loop in 2D,  as also considered in \cite{Cipolletta:2019:arXiv, Moesta2014,beckwith2011second,gardiner2005unsplit,stone2008athena}. This case considers a magnetized circular field loop propagating with a constant velocity in a surrounding non-magnetized ambient medium within a 2D periodic grid. The initial magnetic field is set by the following analytical prescription
\begin{equation} \label{LoopAdvB}
	B^x, \ B^y = 
	\begin{cases} 
	&-A_\mathrm{loop}y/r, \ A_\mathrm{loop}x/r \quad for \ \ r<R_\mathrm{loop} \\
	&	\qquad \qquad \quad \qquad \quad \ \ 0 \quad for \ \ r\geq R_\mathrm{loop}
	\end{cases}
\end{equation}
where $r = \sqrt{x^2 + y^2}$ is the radial coordinate, $R_\mathrm{loop}$ is the loop radius, $A_\mathrm{loop}$ sets the amplitude of the magnetic field, and $B^z$ is set to zero. The corresponding vector potential prescription from which Equation (\ref{LoopAdvB}) can be obtained is given by $\vec{A}(r) = (0,0,\mathrm{max}[0,A_\mathrm{loop}(R_\mathrm{loop}-r)])$ \cite{gardiner2005unsplit}.

As initial data, we set the density and pressure to $\rho=1.0$ and $p=3.0$ respectively throughout the computational domain. For the loop, we set $A_\mathrm{loop}=0.001$ and $R_\mathrm{loop}=0.3$. The fluid 3-velocity is initialized to $v^i=(1/12, 1/24, 1/24)$, keeping a non-zero vertical velocity, i.e., $v^z\neq 0$. The test is performed on a $256\times 256$ grid, where the $x$- and $y$-components span the range [-0.5,0.5]. The CFL factor is 0.25 and the adiabatic index for the ideal EOS is set to $\Gamma=5/3$.  Here, too, we use the MINMOD reconstruction method along with the HLLE flux solver and the RK4 method for time-udpate. 

\begin{figure}[!ht]
	\begin{center}
		\includegraphics[width=0.65\linewidth]{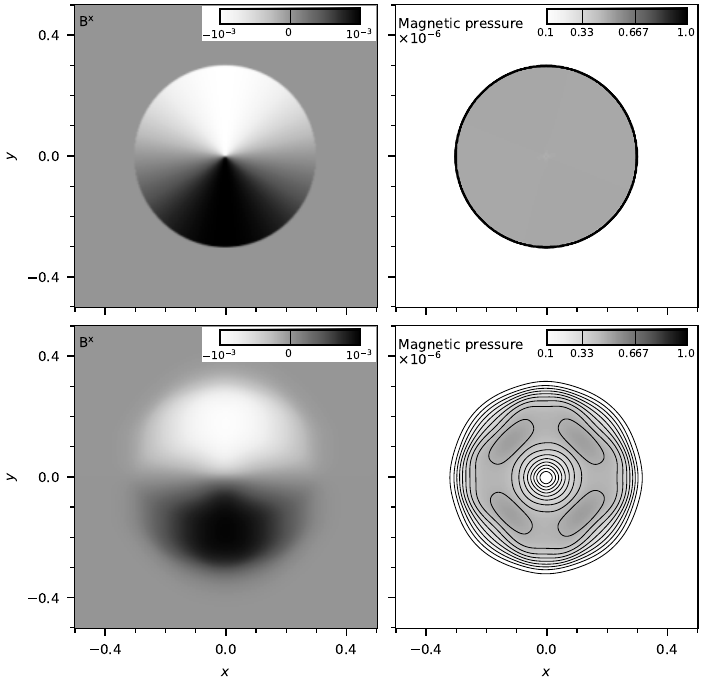}
		\caption{Loop advection test with $v^z=1/24$. Left and right columns represent the x-component of the magnetic field $B^x$ and the magnetic pressure $p_\mathrm{mag} = b^2/2$, respectively. The top row shows the initial data for $B^x$ and its corresponding $p_\mathrm{mag}$ at $t=0$ while the bottom row depicts these quantities after one periodic cycle of evolution, i.e., at $t=24$. Our results exhibit good agreement with those reported in \cite{Cipolletta:2019:arXiv}. }
		\label{LoopAdv2D}
	\end{center}
\end{figure}

In Figure~\ref{LoopAdv2D}, the top row depicts the initial configuration of the magnetic loop for the quantities $B^x$ and $p_\mathrm{mag}=b^2/2$ at $t=0$.
After one entire cycle of evolution across the domain, the same quantities are illustrated in the bottom row at $t=24$. A significant loss of magnetic pressure can be noticed due to numerical dissipation after one evolution cycle, as also reported in \cite{Cipolletta:2019:arXiv}. We also perform another simulation using a less dissipative PPM reconstruction, and find similar final outcome profiles as in \cite{Cipolletta:2019:arXiv}.
 
%%%%%%%%%%%%%%%%%%%%%%%%%%%%%%%%%%
\subsection{Kelvin-Helmholtz instability}
\label{KHI}
The Kelvin-Helmholtz instability (KHI) is an instability that develops across fluid interfaces in the presence of a tangential shear flow, resulting in the production of fluid turbulences. Here, we performed the KHI test in 2D, to evaluate the block-structured AMR feature provided via \texttt{CarpetX}. This test adopts the initial setup similar to the one presented in \cite{Springel2010smoothed}, based on which the simulation domain is $[-0.5, +0.5]^2$ on the xy plane with periodic boundary conditions along x-direction. The central region $|y|<0.25$ is initialized with density $\rho=2.0$ and velocity $v^x=0.5$, whereas the surrounding gas is set up with $\rho=1.0$ and velocity $v^x=-0.5$. The initial gas pressure is kept constant throughout the domain with value $p=2.5$. This gives two contact discontinuities or `slip' surfaces at $|y|=0.25$. To induce the instability, we excite a single mode with wavelength equal to half the domain size by perturbing $v^y$, given by the expression
\begin{equation} \label{KHIvy}
v^y = \omega_0 \sin(4\pi x) \Big( e^ {\frac{-(y-0.25)^2}{2\sigma^2}} + e^ {\ \frac{-(y+0.25)^2}{2\sigma^2}} \Big) ,
\end{equation}
where we set the wave amplitude $\omega_0=0.1$, and $\sigma=0.05/\sqrt{2}$. For simplicity, the magnetic field components are set to zero.

Here, we incorporate 3 levels of AMR with $64^2$ cells on the coarsest level. The error threshold for regridding is set to 0.4. At the time of regridding, regions with first derivatives of either of the fluid variables ($\rho, p, \epsilon, v^i$) that surpass this threshold are refined. We evolve with system up to $t=1.5$ with an ideal gas EOS using $\Gamma=5/3$, along with the MINMOD reconstruction method, the HLLE flux solver, and RK4 time-stepping with CFL factor $0.5$. 

\begin{figure}[!ht]
\begin{center}
\includegraphics[width=0.7\linewidth]{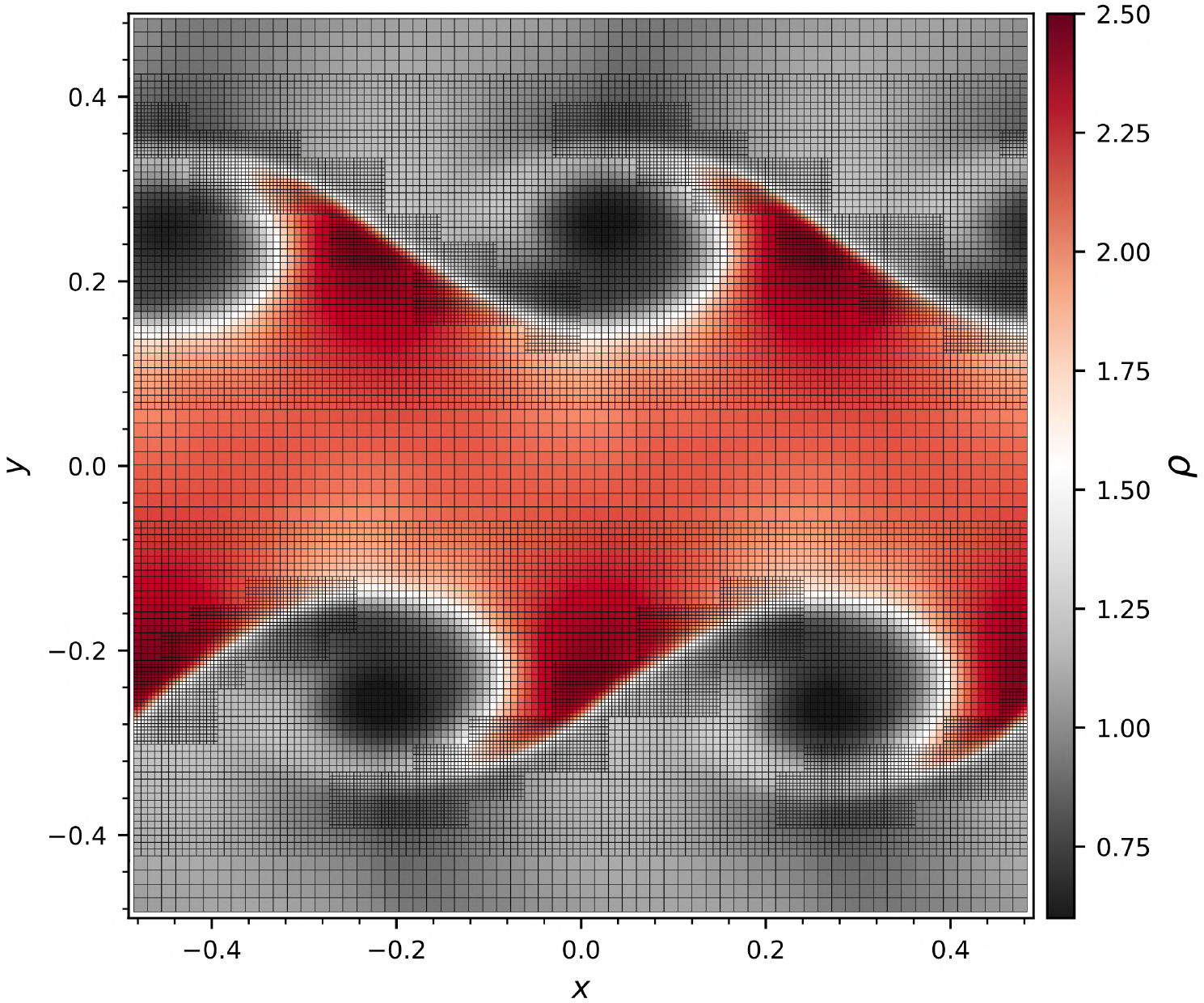}
\caption{Density profile of the Kelvin-Helmholtz instability test at final time=1.5, illustrated along with the grid mesh. The wave-like patterns, that develop at the shear layers, are well-captured by the finest mesh refinement blocks.}
\label{KHI2D}
\end{center}
\end{figure}

In Figure~\ref{KHI2D}, the density profile at end is illustrated along with the block-structured AMR mesh. Here, we notice that KHI grows at the slip surfaces, and develops the characteristic wave-like structures. In order to capture the physics at the shear layers, these waves are refined based on our refinement criterion. However, we also note that the refinement criteria are problem-dependent, and can be improved with further experimentation for this case. Our test demonstrates \texttt{AsterX}'s ability to utilize the dynamic block-structured AMR, which could be essential for accurately resolving shocks and instabilities in various astrophysical fluid flows. One such application of our code with block-structured AMR could be to better capture KHI in binary neutron star (NS) merger simulations, by increasing resolution in regions of interest, i.e. in the shear layer where the KHI develops when the two NS cores touch each other, allowing to effectively model magnetic field amplification.
%

%%%%%%%%%%%%%%%%%%%%%%%%%%%%%%%%%%
%\subsection{3D tests}
%\label{3D}
%\todo{Jay}
%%%%%%%%%%%%%%%%%%%%%%%%%%%%%%%%%%
\subsection{TOV star}
\label{tov}

The Tolman-Oppenheimer-Volkoff (TOV) equations \cite{oppenheimer1939jr,tolman1939static} are typically employed to describe static, spherically symmetric stars in general relativity. As the first 3D test to further assess the stability and accuracy of our code, we consider the evolution of a non-rotating stable magnetized TOV configuration in dynamical spacetime. For computing the initial data, we use our module {\tt TOVSolverX}, ported to the new \texttt{CarpetX} infrastructure from an earlier version \cite{etk2023}. The initial TOV configuration is generated using a polytropic EOS with adiabatic index $\Gamma=2.0$, polytropic constant $K=100$, and initial central rest-mass density $\rho=1.28\times10^{-3}$. This setup is then evolved using an ideal fluid EOS with the same $\Gamma$. In addition, we add a magnetic field to the TOV configuration by hand using the analytical prescription of the vector potential $A_\phi$ given by
\begin{equation}\label{magtovAvec}
	A_\phi \equiv A_\mathrm{b} \varpi^2 {\rm max} \left( p - p_\mathrm{cut}, 0 \right)^{n_s} \ ,
\end{equation}
where $\varpi$ is the cylindrical radius, $A_\mathrm{b}$ is a constant, $p_\mathrm{cut}=0.04p_\mathrm{max}$ determines the cutoff when the magnetic field goes to zero inside the star, with $p_\mathrm{max}$ corresponding to the initial maximum gas pressure, and $n_s=2$ sets the degree of differentiability of the magnetic field strength \cite{giacomazzo2011accurate}. The value of $A_b$ is set such that the maximum value of the initial magnetic field strength is set to $\approx 10^{14} \ \mathrm{G}$. This generates a dipole-like magnetic field confined fully inside the star, which is a standard configuration used for such tests. Exploration of complex magnetic field geometries that include higher-order multipoles and also extend beyond the stellar surface (e.g., \cite{ciolfi2013twisted, suvorov2023magnetic}), in long-term high-resolution simulations can be non-trivial and is deferred to future work. The spacetime is evolved via $Z4c$ with constraint damping parameters $\kappa1 = 0.02$ and $\kappa2 = 0.0$, along with the dissipation coefficient of 0.32. In the atmosphere, we set $\rho_{atm}=10^{-11}$.

The tests are run on a box-in-box AMR grid with 6 refinement levels, having the finest level extending from -20 to 20 along  $x$-, $y$- and $z$-directions, and the radii along with grid-spacings are increased by a factor of 2 for the outer refinement levels. We perform simulations with three different resolutions, i.e., low, medium and high resolution with finest grid spacing of $0.5$, $0.25$ and $0.125$ respectively. All test cases are simulated for $8$~ms using the PPM reconstruction method, the HLLE flux solver, and the RK4 method for time stepping with a CFL factor of $0.25$.

\begin{figure}[!ht] 
	\begin{center}
		\includegraphics[width=0.48\linewidth]{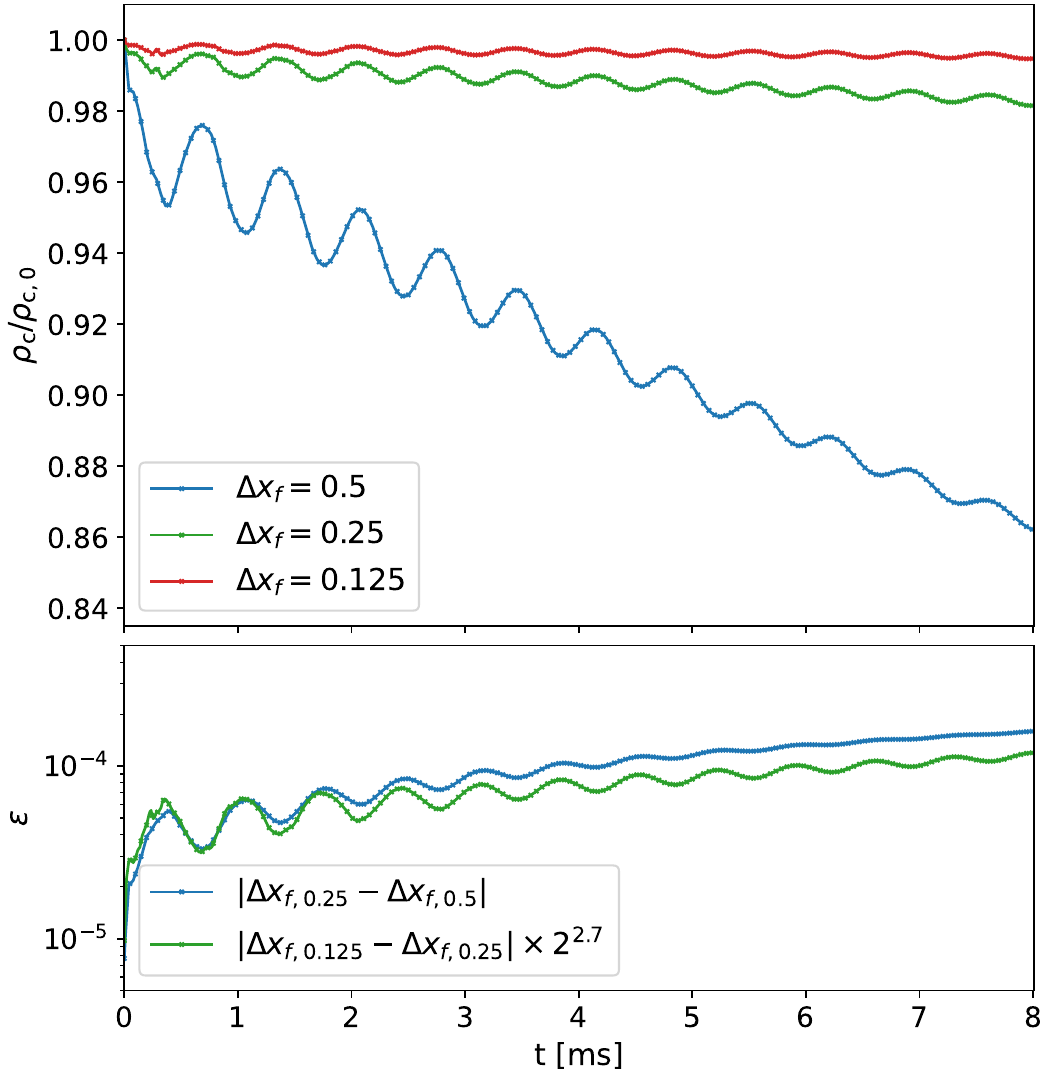}
            \includegraphics[width=0.51\linewidth]{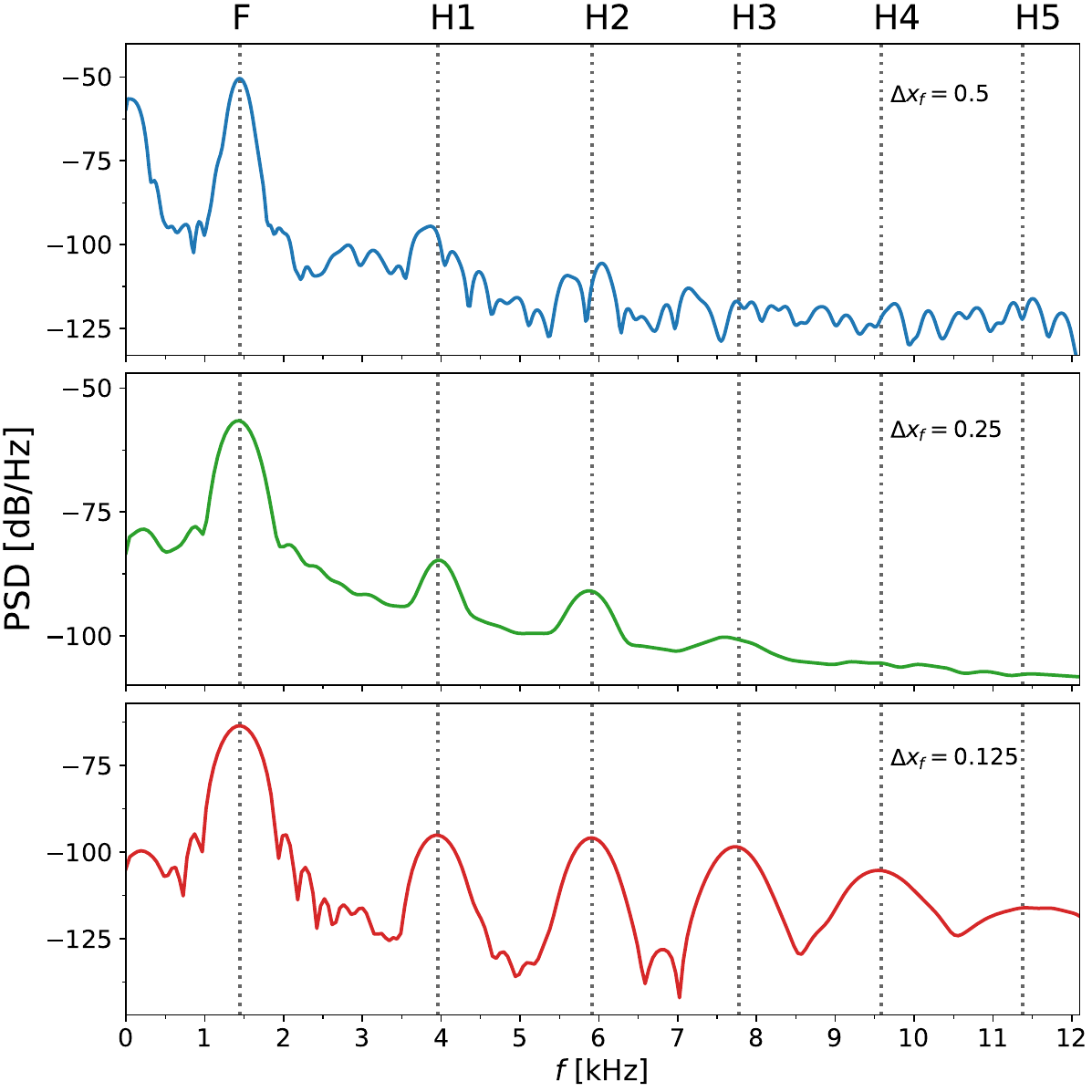}
            \captionof{figure}{Left: The top panel shows the evolution of the normalised central rest-mass density $\rho_\mathrm{c}/\rho_\mathrm{c,0}$ for the different resolution runs. To quantify convergence with increasing resolution,  evolution of the absolute difference $\epsilon$ between the different resolutions are shown in the bottom panel. Right: Top, middle and bottom plots show the power spectrum of the central rest-mass density evolution for the three different resolutions, i.e. $\Delta x_f$ equal to $0.5$, $0.25$ and $0.125$ respectively, compared with peak frequencies of the oscillations (vertical dotted lines) computed from perturbation theory \cite{font2002three}. Here, ${\rm F}$ stands for the fundamental frequency of oscillation whereas ${\rm (H1 \, , H2 \, , H3 \, , H4 \, , H5)}$ represent the higher harmonics.} 
		\label{magtov1D}
	\end{center}
\end{figure}

The top-left panel of Figure \ref{magtov1D} shows the evolution of central rest-mass density $\rho_\mathrm{c}$ for the three simulations with different resolutions. In all cases, the star experiences periodic oscillations induced as a result of the truncation errors present in the initial data, whereas the numerical viscosity of the employed finite differencing (FD) scheme are the main cause of dissipation. The low-resolution simulation experiences both large oscillations as well as heavy dissipation. As we move to higher resolution, the oscillations are reduced to nearly $1\%$, and show convergence of the order $\approx 2.7$, as shown in bottom-left panel of the same figure.

As another validation check, we report oscillation frequencies of the three TOV simulations with different resolutions in the right panel of Figure \ref{magtov1D}. For each simulation, the power spectrum density is computed via fast Fourier transform (FFT) in order to extract the amplitudes and frequencies of the oscillations of the central rest mass density. To compare, in the same figure, we also illustrate the expected peak frequencies of the oscillations taken from \cite{font2002three}, which were obtained independently through a 2D pertubative code. Here, we find that the low resolution case reproduces only the fundamental oscillation mode correctly, but as we increase the resolution, the medium resolution run is able to capture at least two higher harmonic peak frequencies  ${\rm (H1 \, , H2 \,)}$ whereas the high resolution case is able to show matching peak frequencies at least up to fourth harmonic ${\rm (H4)}$. Note that even though we evolve the TOV using ideal gas EOS, which is different from the polytropic EOS used in \cite{font2002three}, the peak frequencies still show a good agreement since ideal gas EOS is expected to produce different results from a polytropic one only in presence of shocks, which appear only near the low-density TOV surface, thus having negligible impact on the oscillations at the core. Our results are also in agreement with the ones reported in the literature \cite{Shankar2023, Cipolletta:2019:arXiv}.

\section{Performance benchmarks}
\label{perf}
To evaluate the performance of \texttt{AsterX}, we conduct scaling tests on OLCF's supercomputer Frontier. Each Frontier compute node is equipped with a 64-core AMD CPU having 512 GB of DDR4 memory, alongside 4 AMD MI250X accelerator units, each featuring 2 Graphic Compute Dies (GCDs). This configuration yields 8 logical GPUs per node, each outfitted with 64 GB of high-bandwidth memory. As noted in \cite{Shankar2023}, the memory available per GPU dictates the problem size, constraining the maximum number of cells per GPU. Furthermore, to maximize GPU utilization, computations should primarily occur within GPU kernels, and CPU-GPU memory transfers should be kept to a minimum.

We perform scaling tests for the following configurations, based on magnetized TOV simulations using an ideal gas EOS for evolution and fourth order Runge-Kutta (RK4) for time-stepping:
\begin{itemize}
\item Setup A: static (Cowling) spacetime + uniform grid
\item Setup B: dynamic (Z4c) spacetime + uniform grid
\item Setup C: dynamic (Z4c) spacetime + 8-level AMR
\end{itemize}
These tests employ PPM reconstruction, an HLLE Riemann solver, Neumann boundary conditions, and 3 ghost zones. For benchmarking purposes only, we refrain from writing any output files and exclude the time taken to set up initial conditions. We quantify our performance in terms of `zone-cycles/s' and `zone-cycles/s/GPU'. Each zone-cycle (ZC) corresponds to the evolution of a single grid cell for one full time-step, encompassing all time-integrator (RK4) sub-steps. Thus, `ZCs/s' provides an estimate of the overall simulation efficiency, while `ZCs/s/GPU' measures the code performance per GPU.

%To be able to compare zone cycle counts between simulations that do not use subcycling and simulations that do, we take the view that subcycling is a performance optimization. Any simulation performed with subcycling could also be run without, using the same amount of memory, with likely a relatively small improvement in accuracy. We define zone cycle counts \emph{as if no subcycling had been used}, i.e. pretending that all grid cells take the same (the smallest) time step size. We agree that this definition is not ideal but we opine that it works best for the purposes of this section. For reference, when using 8 levels of AMR with subcycling, and if every level has the same number of cells, then this over-counts actual zone cycles by about a factor of $6$.

By using subcycling, we reduce the number of ZCs required to reach the same time point by allowing larger time step sizes for coarser levels, compared to the non-subcycling case where all levels evolve using the same time step. As a result, less time is needed to reach the same point.
For an effective comparison, instead of using the wall time required to reach the same time point, we use the total amount of work divided by the wall time taken. Here, we define the zone cycles required by the non-subcycling simulation as the measure of total work.

\subsection{Single GPU benchmarks}
To quantify the performance on a single Frontier GPU (i.e. 1 GCD), we consider a compute load of $240^3$ cells. The results are summarized in Table \ref{tab-single}.  

\begin{table*}
\footnotesize
\begin{center}
\begin{tabular}{@{}ccccc}
\toprule
             & Setup A & Setup B & Setup C & Setup C + subcycling \\
\midrule
levels       & $1$     & $1$     & $8$     & $8$                  \\
cells/level  & $240^3$ & $240^3$ & $120^3$ & $120^3$              \\
cells/GPU    & $240^3$ & $240^3$ & $240^3$ & $240^3$              \\
speed [ZCs/s] & $0.27\!\times\!10^8$ & $0.095\!\times\!10^8$ & $0.072\!\times\!10^8$ & $0.2\!\times\!10^8$ \\
\bottomrule
\end{tabular}
\caption{\label{tab-single} Single-GPU performance tests for our four setups: Setup A (static spacetime + uniform grid), Setup B (dynamical spacetime + uniform grid), Setup C (dynamical spacetime + AMR, without and with subcycling).}
\end{center}
\end{table*}

For Setup A, the simulation runs with $0.46\!\times\!10^8$ ZCs/s when using RK2 integrator, which is comparable to the one reported in \cite{Shankar2023} for a similar setup, which was run however on a single Summit NVIDIA V100 GPU. Whereas, the performance drops to $0.27\!\times\!10^8$ ZCs/s when using RK4 instead. Since RK4 has double the number of sub-steps as compared to RK2, we find the 40\% decrease in run speed to be reasonable. Setup B instead yields a speed of $0.095\!\times\!10^8$ ZCs/s, which indicates that the spacetime solver (Z4c) is roughly twice as expensive as the GRMHD code\footnote{Z4c is hand-written in C++ which makes it difficult to optimize. Other spacetime evolution codes based on automated code generation are in preparation.}. For Setup C, we use $120^3$ cells per level, which effectively gives a compute load of $240^3$ cells in total as well. Without subcycling, we find the performance of $0.072\!\times\!10^8$ ZCs/s, which is about 3.5 and 1.3 times slower than that of Setup A and B respectively. Inclusion of subcycling in time increases the speed to $0.2\!\times\!10^8$ ZCs/s, thus proving to be faster by a factor of about 2.8.

We repeated some of the above tests with the NVIDIA Quadro RTX 5000 GPU available on the Frontera cluster, and encounter an overall reduction in performance by a factor of about 8 for uniform grid cases and about 15 times when using AMR.

We also ran Setup B with the aforementioned problem size on a single Frontier node (8 GPUs), and found the parallel efficiency to decrease to $80\%$ compared to a single GPU performance. This drop is likely due to the additional ghost cell communication as well as decrease in compute load per GPU. The same setup was also run using only CPUs on a single Frontier node (64 CPUs), and we find the CPU run to be about an order of magnitude slower than the GPU one.
 
\subsection{Profiling}
Profiling is a fundamental tool to pinpoint the bottlenecks that make the code slow and inefficient. To identify the most expensive GPU kernels in \texttt{AsterX}, we make use of the \texttt{HPCToolkit} software \cite{adhianto2010hpctoolkit} on Frontier. Specifically, we repeat the simulations for Setup A and B with the compute load of $240^3$ cells on a single GPU, but enabling all the necessary options to output the measurement statistics for \texttt{HPCToolkit}. Here, we also ignore the computational costs for initialization routines.

For Setup A, in which we do not evolve spacetime, we find the most expensive kernel to be \texttt{AsterX\_Fluxes}, which dictates about $70\%$ of the total evolution time. This is expected since this routine involves computations for the reconstruction scheme (PPM) as well as the approximate Riemann solver (HLLE). The next two expensive kernels, i.e. \texttt{AsterX\_SourceTerms} and \texttt{AsterX\_Con2Prim}, which involve calculations of the source terms and the C2P scheme respectively, consume about $11\%$ and $10\%$ of the evolution time. Instead, for Setup B, the spacetime solver kernels prove to be most expensive, as also reported in \cite{Shankar2023}. Here, \texttt{Z4c\_RHS}, which computes the RHSs of the Z4c evolution variables, demands $34\%$, whereas, \texttt{Z4c\_ADM2}, which computes the first time derivative of extrinsic curvature and second time derivatives of gauge quantities, consumes $23\%$ of the total evolution. In this case, \texttt{AsterX\_Fluxes} utilizes $21\%$ whereas computation of $T_{\mu\nu}$ via \texttt{AsterX\_Tmunu} takes $14\%$ of run time. Since the spacetime and the flux solver kernels deem to be most costly, our future efforts will focus on optimizing them.

%%%%%%%%%%%%%%%%%%%%%%%%%%%%%%%%%%
\subsection{Strong scaling}
\label{strong}
Strong scaling measures the efficiency of the parallel performance of the code run with an increasing number of processors while keeping a fixed problem size. Ideally, in this case, increasing the number of GPUs should reduce the time to the solution proportionally, however, achieving perfect linear speedup on GPUs can be challenging due to factors such as communication overhead between GPUs, memory bandwidth limitations, and diminishing returns as the number of GPUs increases.

\begin{figure}[hbt!]
\begin{center}
\includegraphics[width=0.6\linewidth]{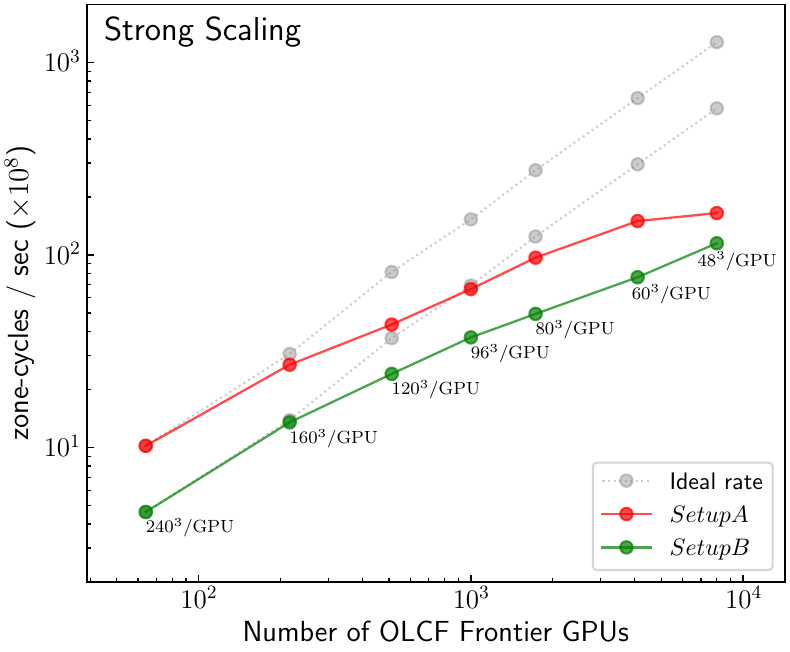}
\captionof{figure}{Results of strong scaling test performed for Setup A (static spacetime + uniform grid) in solid red line, and Setup B (dynamical spacetime + uniform grid) in solid green line. Corresponding ideal rates are shown in light grey dotted lines. Here, we vary the number of GPUs (nodes) from 64 (8) to 8000 (1000) as shown along the x-axis, while keeping a fixed problem size of $960\!\times\!960\!\times\!960$. Performance is quantified in terms of zone-cycles/s, shown along the y-axis. As we increase the number of GPUs, the compute load per GPU varies from $240^3$ cells/GPU to $48^3$ cells/GPU. For 512 GPUs, the strong scaling efficiency $\eta_s$ is maintained at $54\%$ and $65\%$ for Setup A and B, but drops further to $13\%$ and $20\%$ respectively when using 8000 GPUs. One major reason for the decrease in $\eta_s$ stems from the insufficient compute load per GPU, leading to its ineffective usage.}
\label{fig:strongscaling}
\end{center}
\end{figure}

For \texttt{AsterX}, the strong scaling test results are illustrated in Figure \ref{fig:strongscaling}, and details presented in Table \ref{tab2}.
\begin{table*}
\footnotesize
\begin{center}
\begin{tabular}{@{}cccccccc}
  \toprule	
  & & &
  \multicolumn{2}{c}{ Setup A } & 
  \multicolumn{2}{c}{ Setup B } \\ 
  \cmidrule(lr){4-5} \cmidrule(lr){6-7}  
  Nodes & GPUs & cells/GPU & ZCs/s & $\eta_s$ & ZCs/s & $\eta_s$ \\
  \midrule
   8 & 64 & $240^3$ & 
   $10.0 \times 10^8$ & 1.0 & 
   $4.6 \times 10^8$ & 1.0 \\

   27 & 216 & $160^3$ & 
   $26.9  \times 10^8$ & 0.79 & 
   $13.5 \times 10^8$ & 0.86\\

   64 & 512 & $120^3$ & 
   $43.5 \times 10^8$ & 0.54 & 
   $24.1 \times 10^8$ & 0.65\\

   125 & 1000 & $96^3$ & 
   $66.5 \times 10^8$ & 0.42 & 
   $37.3 \times 10^8$ & 0.51\\

   216 & 1728 & $80^3$ & 
   $96.8 \times 10^8$ & 0.36 & 
   $49.4 \times 10^8$ & 0.40 \\

   512 & 4096 & $60^3$ & 
   $150.2 \times 10^8$ & 0.23 & 
   $76.6 \times 10^8$ & 0.26 \\

   1000 & 8000 & $48^3$ & 
   $164.6 \times 10^8$ & 0.13 & 
   $114.8 \times 10^8$ & 0.20 \\
   
  \bottomrule
  \end{tabular}\\
  \caption{\label{tab2} Strong scaling test results for Setup A (static spacetime + uniform grid) and Setup B (dynamical spacetime + uniform grid) with a fixed problem size of $960\!\!\times\!\!960\!\!\times\!\!960$. The strong scaling efficiency $\eta_s$ is normalized by the 8 node performance.}
  \end{center}
\end{table*}
On uniform grid, we consider a problem size of $960\!\!\times\!\!960\!\!\times\!\!960$, and vary the computational load per GPU from $240^3$ cells/GPU to $48^3$ cells/GPU with increasing number of nodes, similar to the study reported in \cite{Shankar2023}. Due to limited computational resources, we perform each test up to 512 iterations. Our parallel efficiency for strong scaling $\eta_s$ is normalized based on the results obtained for 8 nodes (64 GPUs), for which we have $240^3$ cells/GPU. Here, we find that Setup A and B run with about $10\!\times\!10^8$ and $4.6\!\times\!10^8$ ZCs/s respectively. With an about $\times\!16$ increase in node count, $\eta_s$ drops to $42\%$ and $51\%$ for Setup A and B, and decreases further to $13\%$ and $20\%$ respectively with a $\times\!125$ node count increase. This decline in parallel efficiency can be explained by the significantly larger ratio of throughput and memory bandwidth to problem size on GPUs. In other words, we are not efficiently using the GPUs since the problem size per GPU becomes too small, and for the chosen simulation setup, operating below a compute load of about $96^3$ cells/GPU results in significant inefficiency. We also find that for a given number of nodes, the performance in terms of ZCs/s is roughly twice as faster for Setup A which assumes static spacetime, than that of Setup B, which involves evolving the spacetime equations as well. In contrast, the parallel efficiency $\eta_s$ is higher by roughly $10\%$ to $20\%$ for Setup B when compared to that of Setup A. This is expected since the work load per GPU for Setup B is larger, in terms of solving the number of equations, as compared to Setup A.

For Setup C, which includes 8 levels of mesh refinement, we restrict the problem size to $480\!\!\times\!\!480\!\!\times\!\!480$ per level, and vary the computational load from $120^3$ cells/GPU/level to $24^3$ cells/GPU/level. In this case, we find that the number of ZCs/s decreases at most by a factor of 2 as compared to the results of Setup B, and the parallel efficiency goes below $10\%$ for a 1000 node run. In particular for the case without subcycling there is a known inefficiency in our code: Each level is independently distributed across all GPUs, which leads to much smaller per-GPU block sizes than if each GPU handled only blocks from a single level.

Overall, we find our results to be similar to the ones reported in \cite{Shankar2023}.

%As we continue to add more complicated physics to \texttt{AsterX}, in terms of reading tabulated EOSs, neutrino/photon radiation transport, as well as higher order schemes, we expect the overall count of ZC/s to decrease.

%%%%%%%%%%%%%%%%%%%%%%%%%%%%%%%%%%
\subsection{Weak scaling}
\label{weak}
Weak scaling involves measuring the variation in the solution time with the number of GPUs, while maintaining a constant computational load per GPU. Ideally, for a perfect weak scaling, the solution time should remain constant, however, in practice, the parallel performance is affected by a number of factors stated earlier, such as inter-GPU and inter-node communication overheads, limited communication bandwidth, load imbalance, and synchronization overheads.

\begin{figure}[hbt!]
\begin{center}
\includegraphics[width=0.6\linewidth]{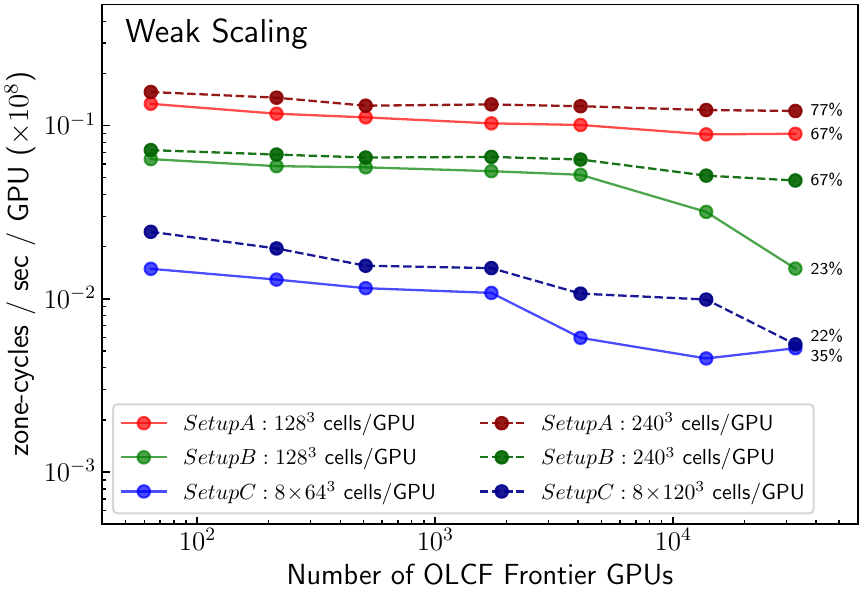}
\captionof{figure}{Results of weak scaling test performed for Setup A (static spacetime + uniform grid), Setup B (dynamical spacetime + uniform grid) and Setup C (dynamical spacetime + 8 levels AMR). Number of GPUs (nodes) is varied from 64 (8) to 32768 (4096) along the x-axis, while keeping a fixed compute load of $128^3$ cells/GPU and $240^3$ cells/GPU for Setup A and B, and effectively for Setup C as well. Along y-axis, we show performance in terms of zone-cycles/s/GPU. In general, Setup A runs are about 2-2.5 times faster than Setup B, and about 6-22 times faster than Setup C. Overall, the runs with higher compute load perform better, for which, with 4096 GPUs, the weak scaling efficiency $\eta_w$ remains at $83\%$ and $88\%$ for Setup A and B, but drops further to $77\%$ and $67\%$ respectively when using 32768 GPUs. For Setup C instead, $\eta_w$ decreases to $62\%$ for the run with 1728 GPUs, and further down to $22\%$ for 32768 GPUs.}
\label{fig:weakscaling}
\end{center}
\end{figure}

\begin{figure}[hbt!]
\begin{center}
\includegraphics[width=0.6\linewidth]{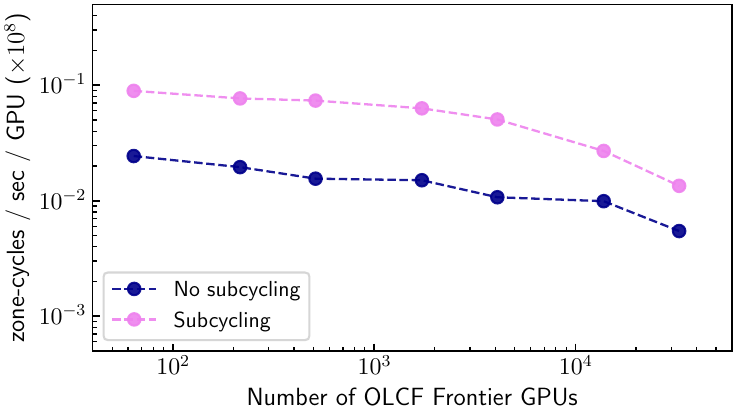}
\captionof{figure}{Comparison of weak scaling results for Setup C (dynamical spacetime + 8 levels AMR) without and with subcycling in time. An overall gain of factor 2.5 to 4.5 is obtained in performance when using subcycling, as expected since larger time steps are employed for coarser refinement levels, whereas runs without subcycling use the same time step size for all levels.}
\label{fig:weakscalingsub}
\end{center}
\end{figure}

Our weak scaling results are illustrated in Figures \ref{fig:weakscaling} and \ref{fig:weakscalingsub}.
\begin{table*}
\footnotesize
\begin{center}
\begin{tabular}{cccccccc}
  \toprule	
  & & &
  \multicolumn{2}{c}{ Setup A } & 
  \multicolumn{2}{c}{ Setup B }  \\ 
  \cmidrule(lr){4-5} \cmidrule(lr){6-7}  
  Nodes & GPUs & Grid & ZCs/s/GPU & $\eta_w$ & ZCs/s/GPU & $\eta_w$  \\
  \midrule
   8 & 64 & $960\!\times\!960\!\times\!960$ & 
   $0.157 \times 10^8$ & 1.0 & 
   $0.072 \times 10^8$ & 1.0  \\

   27 & 216 & $1440\!\times\!1440\!\times\!1440$ & 
   $0.145  \times 10^8$ & 0.92 & 
   $0.068 \times 10^8$ & 0.94  \\

   64 & 512 & $1920\!\times\!1920\!\times\!1920$ & 
   $0.130 \times 10^8$ & 0.83 & 
   $0.065 \times 10^8$ & 0.91  \\

   216 & 1728 & $2880\!\times\!2880\!\times\!2880$ & 
   $0.133 \times 10^8$ & 0.85 & 
   $0.066 \times 10^8$ & 0.91 \\

   512 & 4096 & $3840\!\times\!3840\!\times\!3840$ & 
   $0.129 \times 10^8$ & 0.83 & 
   $0.064 \times 10^8$ & 0.88 \\

   1728 & 13824 & $5760\!\times\!5760\!\times\!5760$ & 
   $0.123 \times 10^8$ & 0.79 & 
   $0.051 \times 10^8$ & 0.71  \\

   4096 & 32768 & $7680\!\times\!7680\!\times\!7680$ & 
   $0.121 \times 10^8$ & 0.77 & 
   $0.048 \times 10^8$ & 0.67 \\
   
  \bottomrule
  \end{tabular}\\
  \caption{\label{tab3} Weak scaling test results for Setup A (static spacetime + uniform grid) and Setup B (dynamical spacetime + uniform grid) with a fixed computational load of $240^3$ cells/GPU. Here, the weak scaling efficiency $\eta_w$ is normalized by the 8 node performance.}
  \end{center}
\end{table*}
\begin{table*}
\footnotesize
\begin{center}
\begin{tabular}{ccccccc}
  \toprule	
  & & & 
  \multicolumn{2}{c}{Setup C} &  
  \multicolumn{2}{c}{Setup C + subcycling} \\ 
  \cmidrule(lr){4-5} \cmidrule(lr){6-7} 
  Nodes & GPUs & Grid & ZCs/s/GPU & $\eta_w$ & ZCs/s/GPU & $\eta_w$  \\
  \midrule
   8 & 64 & $8\!\times\!(480\!\times\!480\!\times\!480)$ & 
   $0.0244 \times 10^8$ & 1.0 & 
   $0.0893 \times 10^8$ & 1.0 \\

   27 & 216 & $8\!\times\!(720\!\times\!720\!\times\!720)$ & 
   $0.0195 \times 10^8$ & 0.8 &
   $0.0767 \times 10^8$ & 0.86 \\

   64 & 512 & $8\!\times\!(960\!\times\!960\!\times\!960)$ & 
   $0.0155 \times 10^8$ & 0.64 & 
   $0.0736 \times 10^8$ & 0.83 \\

   216 & 1728 & $8\!\times\!(1440\!\times\!1440\!\times\!1440)$ & 
   $0.0150 \times 10^8$ & 0.62 & 
   $0.0629 \times 10^8$ & 0.70 \\

   512 & 4096 & $8\!\times\!(1920\!\times\!1920\!\times\!1920)$ & 
   $0.0107 \times 10^8$ & 0.44 &
   $0.0506 \times 10^8$ & 0.57 \\

   1728 & 13824 & $8\!\times\!(2880\!\times\!2880\!\times\!2880)$ &  
   $0.0099 \times 10^8$ & 0.41 & 
   $0.0270 \times 10^8$ & 0.30 \\

   4096 & 32768 & $8\!\times\!(3840\!\times\!3840\!\times\!3840)$ & 
   $0.0055 \times 10^8$ & 0.22 & 
   $0.0135 \times 10^8$ & 0.15 \\
   
  \bottomrule
  \end{tabular}\\
  \caption{\label{tab4} Weak scaling test results for Setup C (dynamical spacetime + 8 levels AMR) with a fixed computational load of $120^3$ cells/GPU/level without and with subcycling in time. Here too, the weak scaling efficiency $\eta_w$ is normalized by the 8 node performance.}
  \end{center}
\end{table*}
For each of the Setups A and B, we consider two cases having a compute load of $128^3$ cells/GPU and $240^3$ cells/GPU, and vary the number of nodes (GPUs) from 8 (64) to 4096 (32768). For Setup C instead, which involves 8 levels of AMR, we consider cases with $64^3$ cells/GPU/level and $120^3$ cells/GPU/level. Summing up the compute load for all 8 levels effectively results in $128^3$ cells/GPU and $240^3$ cells/GPU for Setup C. Results for the case with higher compute load ($240^3$ cells/GPU) are presented in Table \ref{tab3} and \ref{tab4}. Here too, the weak scaling efficiency $\eta_w$ is normalized by the 8-node performance. 

For 8 nodes with $240^3$ cells/GPU, we find the number of ZCs/s/GPU to be about $0.16\!\times\!10^8$ and $0.072\!\times\!10^8$ for Setups A and B respectively. In general, Setup A runs about 2 to 2.5 times faster than Setup B. The weak scaling efficiency $\eta_w$ remains as high as $85\%$ and $91\%$ upto 216 nodes (1728 GPUs) but decreases to $77\%$ and $67\%$ for 4096 nodes (32768 GPUs) for Setups A and B respectively. 
As mentioned earlier, one reason for the decrease in performance is related to the communication overhead. The box-shaped grid structure used for the scaling test distributes the workload per GPU in cubical blocks, where the ghost cell communication scales with the number of blocks as $\sim n^3$. This, in turn, is limited by the communication bandwidth, as also pointed out in \cite{Shankar2023}. 

While the overheads could be minimized by adopting efficient schemes that would require less number of ghost zones, such as Discontinuous Garlerkin methods \cite{Miller2016} and compact finite differencing \cite{Daszuta2024}, increasing the communication bandwidth instead requires improving the cluster infrastructure. Another strategy would be to improve the computation-communication overlap. As noted in \cite{Shankar2023}, the current version of \texttt{CarpetX} does not launch the next GPU kernel until all the ghost-zones from the previous kernel are filled. The ability to launch subsequent kernels that may not immediately require ghost-zone values of the previous kernel while they are being filled, can increase the computation-communication overlap, a feature that is under development in \texttt{CarpetX}.

When considering AMR with $120^3$ cells/GPU/level, we compare performance with and without the activation of subcycling in time. We notice that without subcycling, the 8-node performance goes down to $0.0244\!\times\!10^8$ ZCs/s/GPU, and overall, we notice it to be about 6 to 22 times slower than Setup A (see Figure \ref{fig:weakscaling}). Whereas with subcycling, for 8 nodes, we get $0.0893\!\times\!10^8$ ZCs/s/GPU, and in general, the performance is quantitatively similar to Setup B upto 512 nodes. Overall, we find that subcycling boosts the performance by a factor of about 2.5 to 4.5 when compared to the results without subcycling (see Figure \ref{fig:weakscalingsub}). This is expected since time steps are doubled as we move from finer to coarser grids when using subcycling, whereas a uniform time step is used for all levels for runs without subcycling. In terms of parallel efficiency, $\eta_w$ reduces to $62\%$ ($70\%$) without  (with) subcycling when using 216 nodes (1728 GPUs), and further decreases to $41\%$ ($30\%$) and $22\%$ ($15\%$) for 1728 nodes (13824 GPUs) and 4096 nodes (32768 GPUs) respectively. Here too, we find our results without subcycling to be consistent with the ones reported in \cite{Shankar2023}. 

The drop in ZCs/s/GPU as well as $\eta_w$ can be attributed to reasons such as prolongation/restriction operations for interpolations to finer/coarser grids that demand additional computations and communication, as also highlighted in \cite{Shankar2023}. Also, the existing \texttt{CarpetX} infrastructure enforces each GPU to work on one or more blocks on each refinement level, resulting in serial calculations for the blocks assigned to each GPU. To make this handling efficient, one could let each GPU work on one or more blocks from any given level, thus enhancing weak scaling.

%%%%%%%%%%%%%%%%%%%%%%%%%%%%%%%%%%%%%%%%%%%%%%%%%%%%%%%%%%%%%%%%%%%%%%%%
\section{Conclusions and outlook}
\label{concl}

In this paper, we have presented a new three-dimensional GPU accelerated open-source numerical code \texttt{AsterX} \cite{AsterXGitHub} that solves the full set of GRMHD equations in Cartesian coordinates on a dynamical background. The code is based on \texttt{CarpetX}, the new driver for the Einstein Toolkit which, in turn, is built on top of \texttt{AMReX}, a software machinery designed for massively parallel, block-structured adaptive mesh refinement (AMR) applications. \texttt{AsterX} has been written from scratch in C++ but has adopted several algorithms from \texttt{Spritz} \cite{cipolletta2020spritz}, and the \texttt{RePrimAnd} library \cite{Kastaun2021}, while also benefitting from other publicly available codes such as \texttt{WhiskyTHC} \cite{radice2014high}, \texttt{GRHydro} \cite{Moesta2014} and \texttt{IllinoisGRMHD} \cite{etienne2015illinoisgrmhd}.

%We have described the relevant equations, especially those pertaining to electromagnetic field evolution, and discussed the numerical schemes implemented in the code. 
In \texttt{AsterX}, the employed GRMHD equations are based on the flux-conservative Valencia formulation, and the divergence-free character of magnetic fields is guaranteed by evolving the staggered vector potential. To accurately handle discontinuities, high resolution shock capturing schemes, enabled via reconstruction methods TVD MINMOD and PPM along with Lax-Friedrichs and HLLE flux solvers, have also been utilized. Our primary choice for primitive variable recovery is based on the 2D Noble scheme with 1D Palenzuela routine set as back up. 

\texttt{AsterX} has been validated via a series of classical tests in special and general relativity. We demonstrated the code's ability to accurately reproduce exact solutions by simulating 1D MHD shock tube problems. Whereas, our convergence study illustrates the code to be at least second order accurate. A number of special relativistic MHD tests in 2D were also conducted, namely, cylindrical explosion, magnetic rotor, and loop advection, which proved to be in good agreement with literature. We also simulated the Kelvin-Helmholtz instability in 2D, to examine the handling of block-structured AMR, and confirmed that it performed reliably without issues. For our final test, we evolved a stable magnetized TOV configuration in a dynamical spacetime and successfully obtained peak oscillation frequency modes up to fourth harmonic.

Benchmarking the code on OLCF's Frontier cluster indicated that performance on a single GPU node is at least an order of magnitude faster than the respective CPU node. Strong scaling efficiency is maintained up to 80\% for 1 GPU node when comparing with a single GPU throughput. Whereas, for weak scaling, parallel efficiencies are retained to $67\%-77\%$ for 4096 nodes (32768 GPUs) relative to the 8 node results. Moreover, when employing AMR, use of subcycling provided a significant gain by a factor 2.5-4.5 in performance. Profiling the code also helped us to identify the bottlenecks which will be optimized in the near future.

At present, the code supports different analytical equations of state, and is currently being extended to deal with tabulated microphysical EOSs. The primitive variable recovery algorithm of the \texttt{RePrimAnd} library \cite{Kastaun2021} has also been ported and is undergoing testing, and we plan to add other routines of \cite{Cerda-Duran:2007:169} and \cite{Newman:2014}. In order to obtain higher order convergence, WENO-Z and MP5 reconstruction schemes have already been implemented and are being tested. Next, we intend to include support for advanced Riemann solver HLLD together with upwind CT-HLLD scheme \cite{mignone2021systematic, Kiuchi2022}. For modelling neutrino transport, we have also started looking into porting the moment-based M1 scheme \cite{radice2022new} which would be coupled with the Monte-Carlo scheme, to closely follow the approach of \cite{Izquierdo2024}. These developments will enable us to accurately model astrophysical systems such as core-collapse supernovae and binary neutron star mergers, and will be presented in our next paper. 

Meanwhile, we have already implemented a photon radiation leakage scheme in \texttt{AsterX}, which is currently being tested via black-hole accretion disk simulations and will be presented in another paper. This scheme will then be utilized in our production simulations to study gas dynamics in supermassive binary black-hole mergers, allowing us to reach unprecedented time and length scales via exascale computing.

\hfill\\
\section*{Acknowledgments}
The authors would like to thank Michail Chabanov, Samuel Cupp, Zachariah Etienne, Vassilios Mewes, Phillip Moesta, Swapnil Shankar, Samuel Tootle, and Weiqun Zhang for insightful discussions. We are also grateful to Alexandru Dima, Cheng-Hsin Cheng, Michail Chabanov, and Jake Doherty for providing the \texttt{NewRadX} module. Moreover, we would like to thank the OLCF staff for their support during the Frontier Virtual Hackathon 2023.

We gratefully acknowledge the National Science Foundation (NSF) for financial support from Grants No. PHY-2110338, No. OAC-2004044/1550436/2004157/2004879, No. AST-2009330, No. OAC-1811228, No. OAC-2031744, No. OAC-2310548, No. OAC-2005572, No. OAC-OAC-2103680, and No. PHY1912632, as well as NASA for financial support from NASA TCAN Grants No. 80NSSC18K1488 and No. 80NSSC24K0100 to RIT. Research at Perimeter Institute is supported in part by the Government of Canada through the Department of Innovation, Science and Economic Development and by the Province of Ontario through the Ministry of Colleges and Universities. R.~Ciolfi and B.~Giacomazzo acknowledge support from the European Union under NextGenerationEU, PRIN 2022 Prot. n. 2022KX2Z3B.

This research used resources of the Oak Ridge Leadership Computing Facility provided by Director's Discretionary program on Frontier supercomputer (Grant No. AST182) at the Oak Ridge National Laboratory, which is supported by the Office of Science of the U.S. Department of Energy under Contract No. DE-AC05-00OR22725, and TACC’s Frontera supercomputer allocations (Grants No. PHY20010 and No. AST-20021). Additional resources were provided by RIT’s BlueSky and Green Pairie and Lagoon clusters acquired with NSF Grants No. PHY-2018420, No. PHY-0722703, No. PHY-1229173, and No. PHY1726215. Access to the Deep Bayou cluster was provided by Louisiana State University's HPC group.

%\appendix
%\section{Anythere here?}
%Appendix A..

%%%%%%%%%%%%%%%%%%%%%%%%%%%%%%%%%%%%%%%%%%%%%%%%%%%%%%%%%%%%%%%%%%%%%%%%
%References 
\section*{References}
\bibliographystyle{unsrt}
%\bibliographystyle{plain} % Plain referencing style
%\bibliography{references,biblio}
\bibliography{AsterX}
%%%%%%%%%%%%%%%%%%%%%%%%%%%%%%%%%%%%%%%%%%%%%%%%%%%%%%%%%%%%%%%%%%%%%%%%
\end{document}